\newcommand{\docstyle}{1} %0=elsevier or 1=ieee
\ifnum\docstyle=0
\documentclass[final,5p,times,twocolumn]{elsarticle}
\else
\documentclass[10pt,journal,compsoc]{IEEEtran}
\fi

\usepackage{url}
\usepackage[printonlyused]{acronym}
\usepackage{graphicx}
\usepackage{tabularx}
\usepackage{enumerate}
\usepackage{lipsum}
\usepackage{color}
\usepackage{amsmath,amssymb,amsfonts}
\usepackage{paralist}
\usepackage{threeparttable}
\usepackage{booktabs}
\usepackage{colortbl,hhline}
\usepackage{boldline}
\usepackage{colortbl}
\usepackage{array}
\usepackage{adjustbox}
\usepackage{multirow,tabu}
\usepackage{algorithm,algpseudocode}
\usepackage{tikz,pgfplots}
\usetikzlibrary{arrows}
\usepackage{pgfplotstable}
\usepgfplotslibrary{units,groupplots}
\usepackage{wrapfig}
\usepackage{varwidth,xcolor}
\usepackage{soul}
\usepackage{subcaption}
\usepackage{caption}
\usepackage{stfloats}
\usepackage{blindtext}
\usepackage{pifont}
\usepackage{siunitx}
\usepackage{cleveref}

\newcommand\MarkA{\textsuperscript{$\alpha$}}
\newcommand\MarkB{\textsuperscript{$\beta$}}
\newcommand\MarkC{\textsuperscript{$\theta$}}

\def\eg{e.g.,~}
\def\ie{i.e.,~}

\newcommand{\mytinysize}{\fontsize{6}{7}\selectfont}
\pgfplotsset{
	compat = newest,
	tick label style={font=\sffamily\scriptsize},
	label style={font=\sffamily\scriptsize},
	legend style={font=\sffamily\mytinysize\raggedleft},
	legend cell align=left,
	grid style={dotted,gray}
}

\newcolumntype{?}{!{\vrule width 0.8pt}}

\newlength{\Oldarrayrulewidth}

\begin{document}
\let\WriteBookmarks\relax
\def\floatpagepagefraction{1}
\def\textpagefraction{.001}

\ifnum\docstyle=0
	\journal{Computers \& Security}
\fi

\ifnum\docstyle=0
\title{INTELLECT: Adapting Cyber Threat Detection to Heterogeneous Computing Environments}
\else
\title{INTELLECT: Adapting Cyber Threat Detection to Heterogeneous Computing Environments}
\fi

\author{Simone Magnani\MarkA\MarkB, Liubov Nedoshivina\MarkC, Roberto Doriguzzi-Corin\MarkB, Stefano Braghin\MarkC, Domenico Siracusa\MarkB\\
\MarkA\small\textit{DIBRIS Department, University of Genova, Italy}\\
\MarkB\small\textit{Cybersecurity Centre, Fondazione Bruno Kessler, Italy}\\
\MarkC\small\textit{Security and Privacy, IBM Research Europe, Ireland}}

\graphicspath{{artworks/}{res/}}
\DeclareGraphicsExtensions{.pdf,.jpeg,.png,.jpg,.jpeg}
\pdfinfo{
	/Author (Simone Magnani, Liubov Nedoshivina, Roberto Doriguzzi-Corin, Stefano Braghin, and Domenico Siracusa)
	/Title  (INTELLECT: Adapting Cyber Threat Detection to Heterogeneous Computing Environments)
	/Keywords (Intrusion and Anomaly Detection, Federated Learning, Feature Selection, Model Pruning, Model Fine-Tuning, Catastrophic Forgetting)
}

\acrodef{cnn}[CNN]{Convolutional Neural Network}
\acrodef{ddos}[DDoS]{Distributed Denial of Service}
\acrodef{ebpf}[eBPF]{extended Berkeley Packet Filter}
\acrodef{fnr}[FNR]{False Negative Rate}
\acrodef{fpr}[FPR]{False Positive Rate}
\acrodef{iot}[IoT]{Internet of Things}
\acrodef{lstm}[LSTM]{Long Short Term Memory}
\acrodef{ml}[ML]{Machine Learning}
\acrodef{mlp}[MLP]{Multi-Layer Perceptron}
\acrodef{dl}[DL]{Deep Learning}
\acrodef{cft}[CFT]{Common Features Technique}
\acrodef{llm}[LLM]{Large Language Model}
\acrodef{svm}[SVM]{Support Vector Machine}
\acrodef{nids}[NIDS]{Network Intrusion Detection System}
\acrodef{rnn}[RNN]{Recurrent Neural Network}
\acrodef{zt}[ZT]{Zero-Trust}
\acrodef{hids}[HIDS]{Host Intrusion Detection System}
\acrodef{fl}[FL]{Federated Learning}
\acrodef{vfl}[VFL]{Vertical Federated Learning}
\acrodef{hfl}[HFL]{Horizontal Federated Learning}
\acrodef{ftl}[FTL]{Federated Transfer Learning}
\acrodef{noniid}[non-IID]{non-Independently and Identically Distributed}
\acrodef{fids}[FIDS]{Federated Intrusion Detection System}
\acrodef{he}[HE]{Homomorphic Encryption}
\acrodef{rl}[RL]{Reinforcement Learning}
\acrodef{iiot}[IIoT]{Industrial Internet of Things}
\acrodef{iacs}[IACS]{Industrial Automation and Control System}
\acrodef{gdpr}[GDPR]{General Data Protection Regulation}
\acrodef{iat}[IAT]{Inter Arrival Time}
\acrodef{ids}[IDS]{Intrusion Detection System}
\acrodef{ads}[ADS]{Anomaly Detection System}
\acrodef{iads}[I/ADS]{Intrusion and/or Anomaly Detection System}
\acrodef{tl}[TL]{Transfer Learning}
\acrodef{dp}[DP]{Differential Privacy}
\acrodef{spof}[SPoF]{Single Point of Failure}
\acrodef{pnn}[PNN]{Progressive Neural Network}
\acrodef{nas}[NAS]{Neural Architecture Search}
\acrodef{hpo}[HPO]{Hyperparameter Optimisation}
\acrodef{dnn}[DNN]{Deep Neural Network}
\acrodef{rad}[RAD]{Representation Alignment Dataset}
\acrodef{fedavg}[FedAvg]{Federated Averaging}
\acrodef{fedprox}[FedProx]{Federated Proxing}
\acrodef{cca}[CCA]{Canonical Correlation Analysis}
\acrodef{pd}[PD]{Procrustes Distance}
\acrodef{cka}[CKA]{Centered Kernel Alignment}
\acrodef{simd}[SIMD]{Single Instruction Multiple Data}
\acrodef{rms}[RMS]{Root Mean Square}
\acrodef{rse}[RSE]{Relative Squared Error}

\acrodef{sbe}[SBE]{Sequential Backward Elimination}
\acrodef{sfs}[SFS]{Sequential Forward Selection}
\acrodef{ss}[SS]{Stochastic Search}
\acrodef{es}[ES]{Exhaustive Search}

\acrodef{ca}[CA]{Custom Algorithm}
\acrodef{cap}[CAP]{Custom Algorithm with Perturbation}
\acrodef{rf}[RF]{Random Forest}
\acrodef{fpa}[FPA]{Forest by Penalising Attributes}
\acrodef{pca}[PCA]{Principal Component Analysis}
\acrodef{rfe}[RFE]{Recursive Feature Elimination}

\acrodef{up}[UP]{Unstructural Pruning}
\acrodef{sp}[SP]{Structural Pruning}
\acrodef{drl}[DRL]{Deep Reinforcement Learning}
\acrodef{mse}[MSE]{Mean Squared Error}

\acrodef{name}[INTELLECT]{Integrative Network for Technology-Enabled Learning, Early-warning, and Continuous Training}

\acrodef{dataset17}[CICIDS2017]{Intrusion Detection Evaluation Dataset 2017}
\acrodef{dataset19}[CICIDS2019]{DDoS Evaluation Dataset 2019}

\acrodef{bigml}[BRM]{Base Reference Model}
\acrodef{prunedbigml}[P-BRM]{Pruned BRM}
\acrodef{edgebigml}[E-BRM]{Edge BRM}
\acrodef{edgeprunedbigml}[EP-BRM]{Edge Pruned BRM}

\acrodef{avgtruesoftinferred}[HD]{Hybrid Distillation}
\acrodef{hardtrue}[HT]{Hard True}
\acrodef{hardinferred}[HI]{Hard Inferred}
\acrodef{kd}[KD]{Knowledge Distillation}

\ifnum\docstyle=0
\begin{abstract}
The widespread adoption of cloud computing, edge, and \acl{iot} has increased the attack surface for cyber threats. This is due to the large-scale deployment of often unsecured, heterogeneous devices with varying hardware and software configurations. 
The diversity of these devices attracts a wide array of potential attack methods, making it challenging for individual organizations to have comprehensive knowledge of all possible threats.
In this context, powerful anomaly detection models can be developed by combining data from different parties (\eg organizations, users, devices) using \ac{fl}. \ac{fl} enables the collaborative development of \ac{ml}-based \acp{ids} without requiring the parties to disclose sensitive training data, such as network traffic or sensor readings. 
However, deploying the resulting models can be challenging, as they may require more computational resources than those available on target devices with limited capacity or already allocated for other operations. Training device-specific models is not feasible for an organization because a significant portion of the training data is private to other participants in the \ac{fl} process.

To address these challenges, this paper introduces \acs{name}, a novel solution that integrates feature selection, model pruning, and fine-tuning techniques into a cohesive pipeline for the dynamic adaptation of pre-trained \ac{ml} models and configurations for \acp{ids}. 
Through empirical evaluation, we analyze the benefits of \acs{name}'s approach in tailoring \ac{ml} models to the specific resource constraints of an organization's devices and measure variations in traffic classification accuracy resulting from feature selection, pruning, and fine-tuning operations. Additionally, we demonstrate the advantages of incorporating \acl{kd} techniques while fine-tuning, enabling the \ac{ml} model to consistently adapt to local network patterns while preserving historical knowledge.

\end{abstract}

\begin{keyword}
Intrusion and Anomaly Detection \sep Federated Learning \sep Feature Selection \sep Model Pruning \sep Model Fine-Tuning \sep Catastrophic Forgetting.
\end{keyword}
\maketitle
\else
\maketitle
\begin{abstract}
The widespread adoption of cloud computing, edge, and \acl{iot} has increased the attack surface for cyber threats. This is due to the large-scale deployment of often unsecured, heterogeneous devices with varying hardware and software configurations. 
The diversity of these devices attracts a wide array of potential attack methods, making it challenging for individual organizations to have comprehensive knowledge of all possible threats.
In this context, powerful anomaly detection models can be developed by combining data from different parties (\eg organizations, users, devices) using \ac{fl}. \ac{fl} enables the collaborative development of \ac{ml}-based \acp{ids} without requiring the parties to disclose sensitive training data, such as network traffic or sensor readings. 
However, deploying the resulting models can be challenging, as they may require more computational resources than those available on target devices with limited capacity or already allocated for other operations. Training device-specific models is not feasible for an organization because a significant portion of the training data is private to other participants in the \ac{fl} process.

To address these challenges, this paper introduces \acs{name}, a novel solution that integrates feature selection, model pruning, and fine-tuning techniques into a cohesive pipeline for the dynamic adaptation of pre-trained \ac{ml} models and configurations for \acp{ids}. 
Through empirical evaluation, we analyze the benefits of \acs{name}'s approach in tailoring \ac{ml} models to the specific resource constraints of an organization's devices and measure variations in traffic classification accuracy resulting from feature selection, pruning, and fine-tuning operations. Additionally, we demonstrate the advantages of incorporating \acl{kd} techniques while fine-tuning, enabling the \ac{ml} model to consistently adapt to local network patterns while preserving historical knowledge.

\end{abstract}

\begin{IEEEkeywords}
Intrusion and Anomaly Detection, Federated Learning, Feature Selection, Model Pruning, Model Fine-Tuning, Catastrophic Forgetting.
\end{IEEEkeywords}
\fi

\section{Introduction} \label{sec:intro}

The growth of cloud computing paradigms, edge, and \ac{iot} devices has increased the need for robust defense mechanisms against emerging cyber threats. 
The heterogeneity of such devices and their configurations makes individual organizations exposed to a wide range of cyber threats and zero-day vulnerabilities.
An effective approach to address these emerging security challenges is called \textit{collaborative learning}. Collaborative learning involves training an \ac{ml} model using attack data from multiple sources, such as a consortium of organizations, to obtain a more accurate and efficient \ac{ids}. 

In this context, \acf{fl}~\cite{google} is a collaborative learning technique that allows multiple organizations to jointly develop an \ac{ml}-based \ac{ids} by sharing only the \ac{ml} model's parameters, thereby avoiding the disclosure of potentially sensitive training data. At the end of the process, each organization has access to a copy of this empowered detection model, usually referred to as \textit{global model}, designed with a complex enough architecture to accommodate all data patterns while minimizing misclassification on legitimate traffic.
However, deploying the global model in peripheral devices poses challenges due to the limited computational power, memory, disk space, battery life, and access to data. The model's extended detection capabilities often require large architectures and many input requirements, demanding significant runtime memory and computational capacity. Furthermore, the device's computational capability can also limit the type of traffic observed and preprocessing tasks. For individual organizations, training device-specific models is not a viable option in a \ac{fl} context, as a consistent portion of the training data is private to other participants of the collaborative learning process. As a result, specific adjustments to the global model's architecture and data-gathering requirements are necessary for successful deployment in peripheral devices.

To reduce resource demands, feature selection can be applied to deactivate portions of the data-gathering logic, usually consisting of the less relevant features for the traffic classification, reducing the \ac{ids} footprint for the device~\cite{10.5555/944919.944968}. Another approach involves pruning the global model, simplifying its architecture by removing less relevant parts, therefore creating lightweight models suitable for resource-constrained environments~\cite{Molchanov2016PruningCN}.
However, these operations may degrade the performance of the global model and affect the knowledge acquired during the \ac{fl} rounds. Fine-tuning techniques can help in mitigating the model's performance degradation on the organization's local data but also increase the risk of \textit{catastrophic forgetting} the knowledge obtained from the other participants during the \ac{fl} process. 

So far, our efforts have focused on eliminating less relevant features before training new classifiers~\cite{10175465}, with limited attention given to post-training tasks with limited availability of training data. In our latest work~\cite{10386446}, we presented our framework \acsu{name} (\acl{name}), a methodology to explore the feasible solution space while minimizing the computational and memory footprint of \acp{ids} through feature selection and model pruning.

This paper extends \ac{name} by integrating methods based on \ac{kd} to leverage the knowledge of the full global model (referred to as \ac{bigml} in the rest of this paper) to fine-tune the lighter models produced with feature selection and pruning techniques.
Through an experimental evaluation with open-source datasets, we use \ac{name} to adapt and fine-tune a range of \ac{fl} pre-trained \ac{ids} models to many lighter configurations, simulating the transition from core to the edge.
We investigate the effects of performing feature selection, model pruning, and fine-tuning using the local data available to an \ac{fl} participant, assessing the adapted models’ performance drop with respect to the historical knowledge retained by the \ac{bigml}.
In each test scenario, the framework consistently demonstrated the ability to adapt the given detection models and identify lighter configurations while preserving high accuracy on local data and historical knowledge, enabling deployment in peripheral devices.

The remainder of the paper is organized as follows.
\Cref{sec:related} provides an overview of the state-of-the-art involved. \Cref{sec:methodology} describes the \ac{name} methodology, while \Cref{sec:evaluation} and \Cref{sec:results} present the experimental setup and the obtained results.
Finally, Section~\ref{sec:conclusion} summarises the contribution and highlights directions for future work.
\section{Related Works} \label{sec:related}

In this section, we review the related works on feature selection, model pruning, and model fine-tuning. To the best of our knowledge, no prior work has proposed a solution that integrates these techniques into a single pipeline as \ac{name} does.

\subsection{Feature Selection}

Many techniques to perform feature ranking and selection are highlighted in the literature~\cite{rankingmethods,featsel2,featsel1, featsel3}, discussing their application and challenges related to different problem domains.
These methods include unsupervised methods, such as using correlation or dimensionality reduction techniques like \ac{pca}~\cite{pca} to rank features, and supervised methods that leverage target variables to remove less relevant features. 
The latter can exploit intrinsic properties, such as in \cite{randomforest}, filtering methods that identify the relationship within feature subsets~\cite{pearson, anova, chisquare, mutualinformation}, or wrapper methods such as \ac{rfe} that directly rely on the model performance metrics.

A lightweight and effective feature selection algorithm for~\ac{ids} that combines \ac{rf} and~\mbox{AdaBoost} algorithms is proposed in~\cite{featsel4}. This approach outperforms existing algorithms in terms of detection accuracy and the number of selected features.

The authors in \cite{featsel6} examine the traffic and features of frequently used and recently published datasets for IoT networks. They introduce an ensemble feature selection technique that shows the top and most consistent performance compared to all the classification methods considered.

Automated ensemble feature selection methods are proposed in~\cite{fsadnids,fsids}, where multiple state-of-the-art algorithms are concurrently executed and aggregated. The obtained algorithms outperform individual measures in certain datasets.
 
In \cite{fssupervisednids}, authors carry out a survey and compare many feature selection approaches, using the \ac{dataset17} dataset. They notice that the complexity measured as the time taken to perform the selection tasks increases almost logarithmically with the training set size. Their evaluation framework aims to minimize computation time for each feature ranking algorithm.

\subsection{Model Pruning}

Pruning a neural network means removing the less relevant information from its structures, aiming at decreasing the model memory footprint, computational and storage requirements, while also speeding up the inference process.

There is not a unique globally applicable technique, as each problem and architecture might prefer a different type of pruning.
A few state-of-the-art techniques for pruning neural network models are introduced in \cite{sotapruning,prunecompress,structuredexample,unstructuredexample}.
These methods include:
\textit{(i)} \ac{up}, to remove the less relevant connection in the network; \textit{(ii)} \ac{sp}, which selectively removes a portion of the network like entire layers or neurons; \textit{(iii)} \ac{kd}, designed to train a lightweight \textit{student} model from scratch, possibly with a different architecture, derived from the complex \textit{teacher} network.
Moreover, within the \ac{up} and \ac{sp}, two main strategies exist for pruning network elements: global evaluation of the entire network or local evaluation of each layer independently.

In \cite{finepruning}, the authors focus on pruning the least active neurons susceptible to activation backdoor attacks. They propose fine pruning as a defense against pruning-aware attack mechanisms, which consists of additional rounds of training post-network pruning.

In \cite{scalpel} a \ac{simd}-aware approach to weight and node pruning is proposed, leading to reduced memory footprint and faster predictions. The study reveals that simply removing weights could slow down network prediction due to changes to the low-level parallelism.

For convolutional neural networks, in \cite{pruneconv} is proposed a novel Taylor-expansion-based criterion for pruning, in the context of \ac{tl}. The method proves optimal and better performance compared to traditional kernel weight activation criteria.
Furthermore, in \cite{pruneconv2} the focus shifts to pruning convolutional filters instead of weights, resulting in non-sparse connectivity patterns and decreased computational overhead. Filter importance is computed as the sum of absolute weights (\ie L1 norm), followed by pruning and retraining to improve performance.
Finally, in \cite{prunestruc} the authors highlight the importance of retraining the model after pruning, to better fine-tune its detection capability against the original data. 

\subsection{Model Fine-Tuning}

We now discuss related works in the field of model fine-tuning, the practice of adapting pre-trained \ac{ml} models to specific tasks or domains. This process allows the model to learn task-specific features, improving its performance for specific purposes.

A \ac{rl}-based fine-tuning is proposed in~\cite{10.5555/3600270.3601449}, where authors discuss its usage while fine-tuning large pre-trained models, specifically in the context of \ac{llm}. However, the same technique could be applied in our scenario, where the pre-trained \ac{bigml} model could define the reward function for decisions made by the lighter models, aligning their predictions. Similarly, \ac{rl} can also be used to dynamically adjust the classification threshold, as shown in \cite{rl1}. 

Another widely discussed algorithm is \acf{kd}, involving a teacher model with a certain architecture and parameters to train a student model, which may differ in architecture, neurons, connections, or \ac{ml} algorithm.
In this context, numerous works investigate its applications~\cite{kdsurvey,Romero2014FitNetsHF}. In \cite{kdsimilarity}, authors provide a novel form of \ac{kd} loss to identify similar patterns in the teacher network and distill them in the student model. In \cite{kdpseudo}, the authors try to minimize the knowledge degradation in the student model resulting from distribution shifts in the synthetic data used.
In \cite{Hinton2015DistillingTK}, the authors point out that the distillation process may be sensitive to hyperparameters and might not capture all characteristics of the teacher model.

The research presented in \cite{Lopes2017DataFreeKD} explores data-free knowledge distillation, training the student model without access to the original training data. This approach allows \ac{kd} even in scenarios lacking original data, but information loss during the distillation process, especially without data, could be a limitation for the final performance.
The work presented in \cite{Furlanello2018BornAN} introduces the concept of ``Born-Again'' neural networks, where a simplified student model is trained using knowledge distillation and then fine-tuned on the original task. This method improves the efficiency of the network, but limitations may arise from the need for additional fine-tuning steps and the loss of task-specific information.
In~\cite{Wang2023ACS,Li_2024,8100237}, the authors survey fine-tuning algorithms, while pointing out the advantage of \ac{kd} in addressing catastrophic forgetting.

Another important aspect during fine-tuning rounds is the choice of target labels. In ~\cite{zheng2023learnmodelfinetuningsurvey}, authors survey the literature and analyze the possible combination of methods and labels. The highlighted techniques can be applied also to intrusion and anomaly detection tasks~\cite{nazari}, proving the advantages of soft labels to preserve past knowledge.

To this end, we draw from the presented methods in each field of the literature, respectively leveraging \ac{rfe} for feature selection and implementing \ac{sp} of neurons for model pruning. In addressing the fine-tuning aspect, we analyze and compare conventional learning algorithms, \ac{kd}, and their hybrid combination, while using either hard or soft labels during the learning process.
\section{The INTELLECT Approach} \label{sec:methodology}
\begin{figure}[t]
    \centering
    \includegraphics[width=\columnwidth]{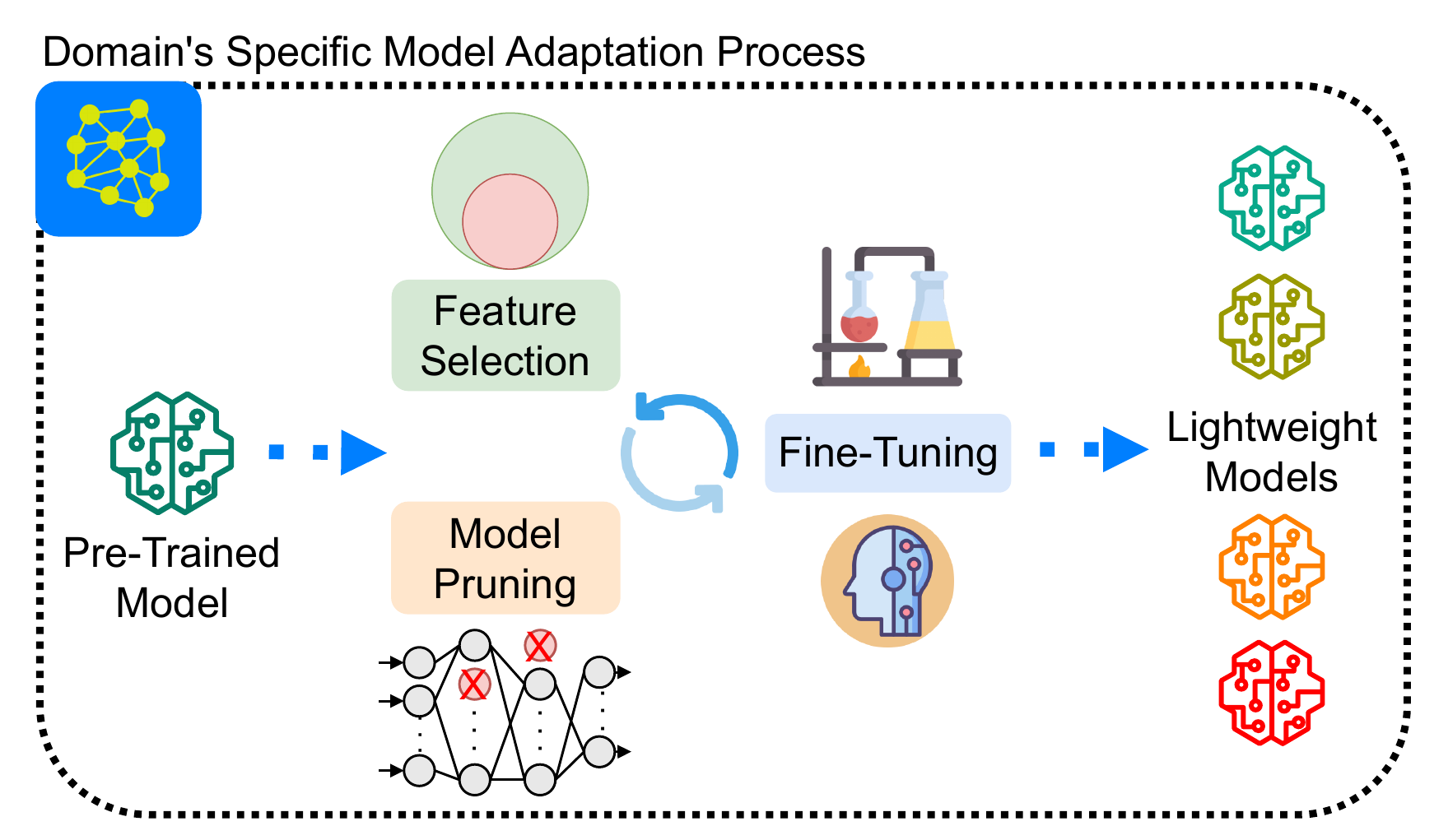}
    \caption{High level architecture of \acs{name}.}
    \label{fig:pipeline}
\end{figure}

\begin{figure}[t]
    \centering
    \includegraphics[width=0.7\columnwidth]{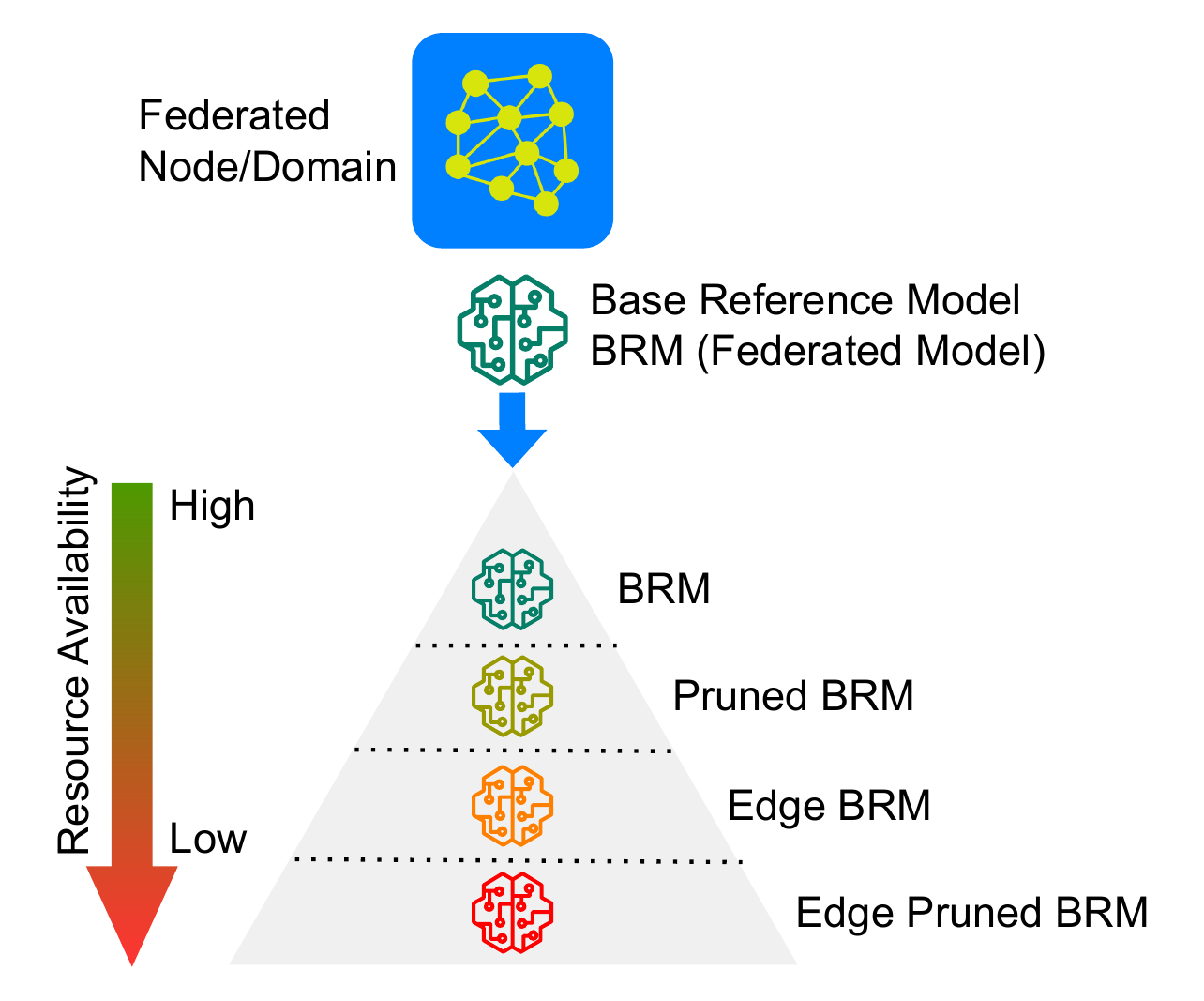}
    \caption{Organization local infrastructure scenario.}
    \label{fig:scenario2}
\end{figure}

In this section, we present the \ac{name} approach, which employs feature selection, model pruning, and fine-tuning techniques (\Cref{fig:pipeline}) to tailor the \ac{fl} global model, referred to as the \acf{bigml} in this paper, to the resource constraints of the target device (\Cref{fig:scenario2}). 
As the model has been trained through \ac{fl}, we assume that the organization cannot access the entire training set during the adaptation process.

\subsection{Base Reference Model Creation} \label{sec:bigmlcreation}

\ac{fl} involves different participants, aiming to share their local insights on anomalies and threats to create a highly accurate global model. However, due to privacy concerns and regulatory constraints, sharing raw private data among participants is uncommon. Instead, \ac{fl} consists of a shared architecture for the detection model, with participants exchanging only the internal status of the model, such as the weights and biases of a neural network, rather than the raw data source.

Depending on the \ac{fl} approach used, these internal model statuses are aggregated to form a new model for the next training rounds. While looking for the optimal architecture, providers can employ different evaluation metrics to measure model performance and loss functions.
During each training iteration, the provider gathers parameters from all participating models and applies an aggregation algorithm to determine the updated global model parameters. This updated state is then distributed to all participants, who adjust their local parameters accordingly. Upon completion of this iterative process, the final version of the global \ac{bigml} is established and a copy of it is available to each participant.

From this moment on, an organization possessing a copy of the \ac{bigml} model may wish to apply feature selection and model pruning methodologies to enable the deployment of lighter versions of the model in resource-constrained devices, such as peripheral edge appliances. In this context, the organization must rely solely on its local data to prune and fine-tune the derived model. This requires a careful analysis of the effects of feature selection and model pruning, including potential performance degradation and loss of historical pattern recognition ability learned during previous \ac{fl} rounds and incorporated into the \ac{bigml}.

\subsection{Feature Selection Process} \label{sec:featsel}

The feature ranking and selection process is a core component in our pipeline, defined by two phases. Initially, feature ranking enables the debugging of the \ac{bigml} model, providing insights into the used input features. Then, feature selection algorithms identify and eliminate redundant or less pertinent features. This optimization allows a more streamlined detection model, reducing both the computational demands of feature extraction and the detection process itself.

Among the algorithms available and discussed in \Cref{sec:related}, we use a perturbation-based algorithm to quantify the significance of each feature for the \ac{bigml} by observing the change in the model's accuracy when the specific feature is perturbed. Our perturbations simulate feature removal, achieved by setting its value to zero. This approach allows us to identify scenarios where certain features can be omitted or not collected, thereby saving computational resources. A high-level procedure adapted from our previous work~\cite{10175465} is illustrated in \Cref{alg:feature-ranking}.

\begin{algorithm}[b!]
    \caption{Feature ranking procedure.}
    \label{alg:feature-ranking}
    \begin{algorithmic}[1]
        \renewcommand{\algorithmicrequire}{\textbf{Input:}}
        \renewcommand{\algorithmicensure}{\textbf{Output:}}
        \Procedure{FeatureRanking}{\ac{ml} model ($w_F$), Current set of features ($F$), Validation set ($X_F$), Labels ($y$)}
            \State $A\gets\emptyset$ \Comment Set of the feature ranking scores
            \State $a_F \gets \Call{Acc}{w_F.predict(X_F,y)}$\label{alg:feature-ranking:accuracy} \Comment Current accuracy
            \For {$f\in F$}
            \State $X_{F\setminus\{f\}}\gets\Call{SetFeatureToZero}{X_F,f}$\label{alg:feature-ranking:zeroed}
            \State $a_{F\setminus\{f\}} \gets \Call{Acc}{w_F.predict(X_{F\setminus\{f\}},y)}$\label{alg:feature-ranking:acczeroed}
            \State $A(f)\gets a_F - a_{F\setminus\{f\}} $
            \EndFor
            \State $f_{min}=\operatorname*{argmin}_{f\in F} A(f)$\label{alg:feature-ranking:argmin}
            \State \textbf{return} $f_{min}$\label{alg:feature-ranking:return} \Comment Return the least important feature
        \EndProcedure
    \end{algorithmic}
\end{algorithm}

\Cref{alg:feature-ranking} is designed to identify the least relevant feature $f_{min}$ within the feature set $F$, for the given \ac{ml} model $w_F$ trained on a labeled training set $X_F$ comprising $F$ features. 
At first, the algorithm computes the model's accuracy using all available features (line \ref{alg:feature-ranking:accuracy}).
Then, for each feature $f\in F$, the values of the corresponding column in $X_F$ are set to zero for all samples (line \ref{alg:feature-ranking:zeroed}), a process commonly termed as ``zeroing out'' the feature $f \in X_F$.
This method aims to quantify the reduction in the model's accuracy when excluding a particular feature $f$ (line \ref{alg:feature-ranking:acczeroed}); the higher the reduction the more important the feature.
After determining the loss for each feature, the algorithm identifies and returns the feature with the minimal loss (lines \ref{alg:feature-ranking:argmin} and \ref{alg:feature-ranking:return}).

With the computed feature ranking, we implement a \ac{sbe} procedure to iteratively identify smaller feature subsets while preserving near-optimal accuracy. In applications like \ac{ids}, a thorough evaluation of feasible features and model complexities adaptable to the target devices is essential to ensure post-deployment system performance integrity. However, challenges may arise due to resource constraints or potential feature unavailability in edge devices, limiting access to consistent information and complexity essential for the model's optimal functioning across diverse infrastructure sections.

To address these challenges, the organization can explore the feasibility and performance implications of reducing the information supplied to the \ac{bigml}. In addition to minimizing data fed into the \ac{bigml}, there is a focus on refining the model by eliminating irrelevant information, aiming to create an optimized version tailored to receive specific inputs, thereby minimizing knowledge loss.
This investigation serves a dual purpose. Firstly, it sheds light on the relevance of each input feature, identifying elements that contribute most to the model's performance. Secondly, it explores the practicality of deploying a detection system using specific feature subsets that may differ across endpoints.

Conducting an exhaustive search to evaluate all potential feature subsets is often impossible due to the exponential complexity of the search space. Therefore, for establishing a feasibility baseline, our approach leverages the \ac{sbe} algorithm, which has demonstrated effective outcomes in less time by iteratively eliminating features based on their impact on model performance.
The computational time to compute the subsets is significantly lower than performing an exhaustive search, as only a fixed number of subsets equal to the total features $F$ exists from the given rank. This reduction provides a practical assessment, showing immediate insights into the impact of using fewer features on the model's performance. 

At this stage of the methodology, the subsets identified by the \ac{sbe} algorithm are highly overlapping: each newly generated subset of size $S_i$ includes all the features from the previous one of size $S_{i+1}$. In practical scenarios, devices with similar capabilities might not generate identical data, or at least not entirely, even if theoretically they could extract the same number of features $F_i$. To address this issue and to diversify the feature subsets, we introduce a \ac{ss} approach. This \ac{ss} aims to identify sub-optimal feature subsets with less overlap, thereby expanding the range of feasible deployment scenarios. The high-level procedure is illustrated in \Cref{alg:subset-search}.

\begin{algorithm}[t!]
    \caption{Subset Search procedure.}
    \label{alg:subset-search}
    \begin{algorithmic}[1]
        \renewcommand{\algorithmicrequire}{\textbf{Input:}}
        \renewcommand{\algorithmicensure}{\textbf{Output:}}
        \Procedure{SubsetSearch}{\ac{ml} model ($w_F$), Features ($F$), Features Rank ($R$), Subset Ratio ($s$), \# Solutions ($n$), Validation set ($X_F$), Labels ($y$)}
            \State $S\gets\emptyset$ \Comment List of the identified subsets
            \State $j\gets \Call{Round}{F.Size \times s}$ \Comment \# Features to draw \label{alg:ss-one}
            \State $P\gets\Call{RankToDistribution}{R}$ \label{alg:ss-two}
            \For {$i=0; i < n; i=i+1$}
            \State $T \gets \Call{WightedDraw}{F, probs=P}$ \label{alg:ss-three}
            \If{ $ T \notin S$}
            \State $a_{F\setminus\{T^\complement\}} \gets \Call{Acc}{w_F.predict(X_{F\setminus\{T^\complement\}},y)}$\label{alg:ss-four}
            \State $S(T) \gets a_{F\setminus\{T^\complement\}}$
            \EndIf
            \EndFor
            \State \textbf{return} $S$
        \EndProcedure
    \end{algorithmic}
\end{algorithm}

The \ac{ss} approach proposed requires three main parameters: the rank of the features previously computed $R$, the ratio of the dimension of the subset to identify $s$, and the number of different solutions to explore $n$.
At first, the ratio is translated into the actual number of features to select (line \ref{alg:ss-one}). Then, the features rank is converted into a distribution of values that sums up to $1$ based on the initial values (line \ref{alg:ss-two}): here, the lower a value the less probability to be chosen is assigned to that feature, allowing the algorithm to increase the probability of finding near-optimal solutions by facilitating the choice of those features rather than the poorly performant ones.
Upon the weighted sampling of $n$ features (line \ref{alg:ss-three}), the algorithm discards the identified solution $T$ in the case is duplicated. Otherwise, it computes the accuracy of the \ac{bigml} with only the identified $T$  features active while removing the others (line \ref{alg:ss-four}). Finally, the algorithm outputs the identified distinct solutions tested against the \ac{bigml}. 

\subsection{Model Pruning}

This phase involves the critical process of refining the \ac{bigml} to streamline its architecture for deployment on resource-limited devices. Pruning a neural network before deployment offers advantages, including reducing the model size by eliminating redundant parameters. This reduction results in a more compact model, often translating to faster inference times and smaller memory footprints.

Our methodology offers flexibility in pruning, enabling the removal of entire layers, individual neurons, or sparse connections within the model's internal structure. The pruning criteria are many, including multiple conditions such as assessing the L1/L2 norm of the targeted item for pruning, computed based on the 2-D matrix that represents the weight magnitudes.
These conditions help to selectively retain the most valuable segments of the network. This selective approach aims to preserve the network's optimal components, thereby minimizing the loss of critical knowledge. By adopting these targeted pruning strategies, our methodology attempts to find a balance between model efficiency and knowledge retention.

Our pruning methodologies support both global and local approaches, enabling application across the entire network or individual layers, respectively. While sparse connection pruning can help address overfitting and facilitate the learning of new patterns, they do not lead to computational or memory efficiency gains. Similarly, during inference, the missing parameters are indicated by zero values rather than truly lighter architecture, providing no computational time savings.
On the other hand, neurons or entire layer removal techniques lead to a more simplified structure. The resulting model requires less disk space, consumes less memory during execution, and needs reduced computational capacity for inference. These pruning methods are particularly suitable for scenarios requiring a lightweight deployment, as for \ac{iot} devices. In this study, we leverage these pruning strategies to enhance the model for efficient deployment on resource-constrained devices. An overview of the procedure is detailed in \Cref{alg:model-pruning}.

\begin{algorithm}[t!]
    \caption{Model Pruning procedure.}
    \label{alg:model-pruning}
    \begin{algorithmic}[1]
        \renewcommand{\algorithmicrequire}{\textbf{Input:}}
        \renewcommand{\algorithmicensure}{\textbf{Output:}}
        \Procedure{ModelPruning}{\ac{ml} model ($w$), Pruning Ratio List ($P$), Validation set ($X$), Labels ($y$)}
            \State $S\gets\emptyset$ \Comment List of the identified ratios
            \State $N\gets\emptyset$ \Comment Reference to Neurons
            \For {$L \in \Call{PrunableLayers}{w_F}$}
            \For {$n \in \Call{l.GetNeurons}{}$}
            \State $N(n)\gets\Call{L1Norm}{n.weight}$ \label{alg:mp-one}
            \EndFor
            \EndFor
            \State $N\gets\Call{SortByValue}{N}$ \label{alg:mp-two}
            \For {$r \in P$}
            \State $m \gets \Call{Copy}{w}$\label{alg:mp-three}
            \State $d \gets N.Size \times r$
            \For {$n \in \Call{Head}{F, d}$}
            \State $\Call{m.Remove}{n}$ \label{alg:mp-four}
            \EndFor
            \State $S(r) \gets \Call{Acc}{m.predict(X, y)}$ \label{alg:mp-five}
            \EndFor
            \State \textbf{return} $S$
        \EndProcedure
    \end{algorithmic}
\end{algorithm}

The procedure requires a list of pruning percentages $P$ to be tested, relative to the initial state and size of the \ac{bigml}. The algorithm first retrieves references to all the neurons located in the area of the network marked as ``prunable'' (line \ref{alg:mp-one}), such as all the hidden layers in our case (input and output are preserved for interpretability). In this step, each neuron is associated with its respective L1-norm and sorted, as in line \ref{alg:mp-two}. For each pruning ratio to be tested, the algorithm creates a copy of the model and applies the pruning algorithm to remove references to the least performing neurons according to the ratio $r$ (line \ref{alg:mp-four}). Finally, the resulting model is tested and its accuracy is stored and associated with the model.

The result of the whole adaptation process is a leaderboard displaying many combinations of pruned models evaluated against the complete set of input features $F$ and the many subsets identified in the previous step (\Cref{sec:featsel}). Each leaderboard entry is labeled with a unique model identifier and subset identifier. The subset identifier enables access to subset-specific statistics, such as the features composing it. In contrast, the model identifier allows retrieval of pruned model-specific statistics, including pruning portion, memory footprint, layer sparsity, and more.

\subsection{Model Fine-Tuning}

We now discuss the final step of our methodology, which involves fine-tuning the identified lighter configurations through feature selection and model pruning. 
In this case, the pre-trained \ac{bigml} has intrinsically high knowledge of many traffic patterns potentially local to each organization previously involved in the \ac{fl} training. While reducing the resource demands of \ac{bigml} to identify a lighter setup, an organization risks degrading the model performance. However, while it is possible to re-establish the knowledge against the traffic data local to the organization, the historical knowledge might vanish, which is in contrast with the premise of adopting \ac{fl} to extend the detection capabilities. In this context, the following research questions arise: \textit{(i)} is it feasible to distill or restore the knowledge loss from the original \ac{bigml} to the derived models? \textit{(ii)} Regardless of this loss, in the event of fine-tuning the model against local data, what would be the effects of catastrophic forgetting? Are there learning algorithms more suitable than others for these scenarios?

We now consider two models: a copy of the original \ac{bigml} that serves as the ``teacher'' model, containing all the initial information on data patterns, and a ``student'' model resulting from prior feature selection and model pruning. While the teacher model guides the learning as an oracle with accurate knowledge of all traffic categories, the student model is the focus of the learning process, where we aim to investigate the effects and assess the quality during the fine-tuning.

\begin{table*}[tb]
    \centering
    \caption{Learning algorithms explanation.}
    \label{tab:learningalgorithms}
    \begin{adjustbox}{width=\textwidth}
    \begin{tabular}{c|c|c}
        \toprule
         Name & Original Labels Required & Description \\
         \midrule
         \acs{hardtrue} & \ding{51} & Uses the values of the original labels ($1/0$).\\
         \acs{avgtruesoftinferred} & \ding{51} & Uses the average of original labels ($1/0$) and soft predictions of the teacher model ($\in [0,1])$.\\
         \acs{hardinferred} & \ding{55} & Uses the values of the labels inferred by the teacher model ($1/0$).\\
         \acs{kd} & \ding{55} & Uses the soft predictions of the teacher model ($\in [0,1])$.\\
         \bottomrule
    \end{tabular}
    \end{adjustbox}
\end{table*}

We identified four different learning algorithms for analysis, as reported in \Cref{tab:learningalgorithms}. First, the \ac{hardtrue} involves conventional training, using the true values of the original data as labels/targets, rounded as $1/0$. This method is a standard learning algorithm commonly used by frameworks~\cite{sklearn}. The \ac{avgtruesoftinferred} approach combines the true labels with the raw prediction of the teacher model. Similarly to \ac{hardtrue}, this algorithm requires the availability of the original true labels to execute. On the other hand, \ac{hardinferred} leverages the labels inferred by the teacher model, rounded at $1/0$ similarly to the \ac{hardtrue}, but without the prerequisite of having the original labels available. Lastly, \ac{kd} relies only on the raw predictions of the teacher model.

These identified algorithms cover most configurations of a potential real-life deployment scenario, considering the availability or lack of labeled data. While the \ac{hardtrue} and \ac{avgtruesoftinferred} algorithms require access to labels, the \ac{hardinferred} and \ac{kd} do not need this information and can be more readily deployed in live scenarios with incoming online data, where the labeling process may not be feasible due to time constraints.

We present in \Cref{eq:kd2} a simplified version of the loss function $L_{KD}'$ implemented in our methodology.

\begin{equation}
    \begin{array}{c}
         L_{KD}' = \alpha q + (1-\alpha)p
    \end{array}
    \label{eq:kd2}
\end{equation}

The adjusted loss $L_{KD}'$ is a weighted combination of the teacher's raw predictions $q$ within the range of $[0,1]$ and the original $0/1$ labels $p$. This formulation does not account for the student predictions' when determining the target labels, but it acts similarly to standard training with predetermined labels. However, this allows for flexibility with the parameter $\alpha$ to weigh the importance of the teacher model to the original data, accommodating scenarios where the true labels may not be available. The \ac{avgtruesoftinferred} leverages this loss function with $\alpha=0.5$, evenly weighting the two values. On the other hand, the \ac{kd} uses $\alpha=1$, ignoring the original true labels to focus on emulating the exact behavior of the teacher model by using soft labels, which capture intrinsic correlations among data.

\subsection{Models Distribution}

Upon identifying the most suitable derived models, they are integrated into the infrastructure to enhance its resilience against cyber threats. The deployment strategy leverages an internal model repository accessible only within the organization's secure environment, with external access restricted for security reasons.

Each model in the catalog is associated with detailed information, including the feature set used, inference runtime overhead, memory and disk usage, and performance metrics. This repository equips system administrators with the ability to select the most suitable devices, enriched with contextual details about the primary application on each device.

The use of a model catalog provides system administrators with the flexibility to use fine-tuning algorithms that may pose challenges in specific contexts, such as on-device training of a model with limited resources. In such scenarios, administrators can fetch the lightweight models from the catalog to a server with enough computational capabilities. Here, they can perform the required training iterations using newly acquired data from the remote device. Then, the updated instance or refined model parameters are reintegrated into the repository. This updated version can then be downloaded to the remote device, favoring adaptive and efficient deployments.

Administrators or devices supporting self-configurable deployment can query the catalog to identify the optimal model for a given scenario, balancing many search parameters. Each device is registered in the system. When deployed, the device starts the authentication process with the server to access the model catalog and obtain its assigned model version. To foresee any issues during the initial device bootstrap, equipping the device with a local model copy is recommended. This ensures synchronization with the catalog version once any connectivity or authentication problems are resolved. Although such situations are rare, all communications between devices and the catalog are encrypted and continuously monitored by administrators.

\section{Experimental Setup} \label{sec:evaluation}

We evaluate \ac{name} using publicly accessible datasets to showcase and emphasize the efficacy of the proposed approach. The code is available\footnote{\url{https://gitlab.fbk.eu/smagnani/intellect}} for replicability.

\subsection{Evaluation Dataset} \label{sec:data}

We use datasets such as \ac{dataset17} and \ac{dataset19}, presented respectively in \cite{Sharafaldin2018TowardGA,8888419}, known for their extensive coverage of many attacks.
Our focus is on anomaly detection rather than specific categorization by type. Therefore, the objective is to develop classifiers for a single binary class ($0/1$), \texttt{BENIGN} and \texttt{MALICIOUS} respectively. The \texttt{MALICIOUS} class includes all the attack categories of the two datasets, namely: \texttt{DDoS}, \texttt{FTP-Patator}, \texttt{DoS slowloris}, \texttt{DrDoS SSDP}, \texttt{LDAP}, \texttt{UDP lag}, \texttt{DrDoS MSSQL}, \texttt{DrDoS NTP}, \texttt{UDP}, \texttt{DrDoS SNMP}, \texttt{DrDoS UDP}, \texttt{DrDoS LDAP}, \texttt{MSSQL}, \texttt{DrDoS DNS}, \texttt{NetBIOS}, \texttt{Portmap}, \texttt{DrDoS NetBIOS}, \texttt{Syn}, \texttt{TFTP}, \texttt{Web Attack Brute Force}, \texttt{PortScan}, \texttt{WebDDoS}, \texttt{Bot}.

We remove non-overlapping features between the two datasets to prevent missing values after merging and discard columns used for sample identification, such as network connection identifiers (\eg 5-tuple with IP addresses, L4 ports, and the carried protocol).
In addition, we remove non-informative features that the classifier would not use or that potentially bias its training process (\ie single-value features, missing records, and more).
Non-numeric features (e.g., HTTP verbs) are encoded into categorical values to align with the input requirements of the neural network. Finally, the features are scaled using the Min-Max normalization technique to constrain them within the $[0,1]$ interval.
The resulting samples are balanced and split into three main sets: the training~(65\%), validation~(15\%), and test sets~(20\%).

\subsection{Methodology} \label{sec:evalmethodology}

For the generation of the \ac{bigml} described in \Cref{sec:bigmlcreation}, we assume that the architecture and hyperparameters are provided -- for example using off-the-shelve pre-trained models -- and that neither the service provider nor each participant in the training has to perform \acl{nas} and \acl{hpo}.
Those decisions are agreed in advance among all organizations involved and the service provider.
The chosen network is an extended version of a \texttt{MlpRegressor} class from the \texttt{scikit-learn} library, for which we developed the pruning algorithms and added support for other learning methods. The network is initialized with $4$ hidden layers, which sums up the input and output layers, with $64$ hidden neurons each. We use the $TanH$ as an activation function and a learning rate of $0.001$. We leverage a batch size of $512$ and set the maximum number of epochs to $100$ with an early stop condition of $25$ epochs with no improvements.
While the evaluations are performed using the entire sets including all the traffic categories, the organization-specific tasks leverage only two balanced portions of the traffic, the \texttt{BENIGN} and \texttt{DDoS}, which were found to be the most distinct in terms of feature distributions. This allows us to measure the effects of fine-tuning the model against local traffic that significantly differs from all the other categories.
The resulting model achieves a near-optimal accuracy equal to $0.99$, which we will use as a baseline in the plots.

We explored all the model pruning ratios as $p_{0.05}$, $p_{0.1}$,~\ldots, up to $p_{0.95}$, allowing the pruning of neurons in the hidden layer of the model based on their L1-norm. We additionally present results while removing internal connections instead of neurons to demonstrate the effectiveness and discuss the different uses of these methods. For the search of feature subsets, we let the algorithms presented in \Cref{sec:featsel} look for $1000$ distinct subsets with a ratio of active feature equal to $s_{0.1}$, $s_{0.2}$, \ldots, up to $s_{0.9}$.

\begin{table}[tb]
    \centering
    \caption{Naming convention of the achievable model given the different availability of information.}
    \label{tab:matrixmodels}
    \begin{tabular}{c|c|c|c}
        \toprule
         Name & Full Model & Full Features & Location \\
         \midrule
         BRM & \ding{51} & \ding{51} & Centralized\\
         P-BRM & \ding{55} & \ding{51} & Centralized\\
         E-BRM & \ding{51} & \ding{55} & Edge\\
         EP-BRM & \ding{55} & \ding{55} & Edge\\
         \bottomrule
    \end{tabular}
\end{table}

Combining the feature subset selection with the model pruning leads to a different resulting model, whose nomenclature is presented in \Cref{tab:matrixmodels}. At the final stage of the \ac{fl} training process, the \ac{bigml} is achieved. When performing only model pruning, we obtain what we define as \ac{prunedbigml}. On the other hand, keeping the original \ac{bigml} but removing features from the input set leads to the \ac{edgebigml}. Finally, combining the feature removal with the model pruning technique, we achieve the defined \ac{edgeprunedbigml}.
In our methodology, the key distinction between centralized and edge model deployment is the potentially limited availability of features once the model is deployed on the device. For edge deployment, we investigate the effects of fine-tuning the models with a similar input where certain features are missing, to better simulate their future operating conditions.

\begin{table}[tb]
    \centering
    \caption{Naming convention of the identified scenarios when deploying a student model in the edge.}
    \label{tab:matrixscenarios}
    \begin{tabular}{c|c|c}
        \toprule
         Name & Teacher has Full Features & Learning Input \\
         \midrule
         Case 1 & \ding{55} & Student\\
         Case 2 & \ding{51} & Student\\
         Case 3 & \ding{51} & Teacher\\
         Case 4 & \ding{51} & Mixed\\
         \bottomrule
    \end{tabular}
\end{table}

For the assessment of model fine-tuning, we evaluate four algorithms outlined in \Cref{tab:learningalgorithms}, using the original \ac{bigml} as a teacher model when required (\eg when using \ac{kd}, \ac{hardinferred}, and \ac{avgtruesoftinferred}). As the target model for fine-tuning, we test the four identified student models in \Cref{tab:matrixmodels}.
In a centralized deployment (\ie \ac{bigml} and \ac{prunedbigml} in \Cref{tab:matrixmodels}) where there is full availability of the entire feature set, we identify a single operational scenario. The teacher model is always a \ac{bigml}, while the student model, which uses the entire feature set as the teacher, is either a replica of \ac{bigml} or a \ac{prunedbigml}. While the former serves as a baseline to measure the effects of fine-tuning algorithms in ideal conditions, with the \ac{prunedbigml} and the following models we evaluate such effects in incrementally less ideal scenarios.
On the other hand, in edge scenarios (\ie \ac{edgebigml} and \ac{edgeprunedbigml} in \Cref{tab:matrixmodels}), we identify and examine multiple nested cases presented in \Cref{tab:matrixscenarios}, which differ in the richness of input features.
\section{Evaluation Results} \label{sec:results}

We now discuss the obtained results from the testbed setup and configurations described in \Cref{sec:evaluation}. 

\subsection{Feature Subset Search}

\begin{figure}[tb]
    \centering
    \includegraphics[width=\columnwidth]{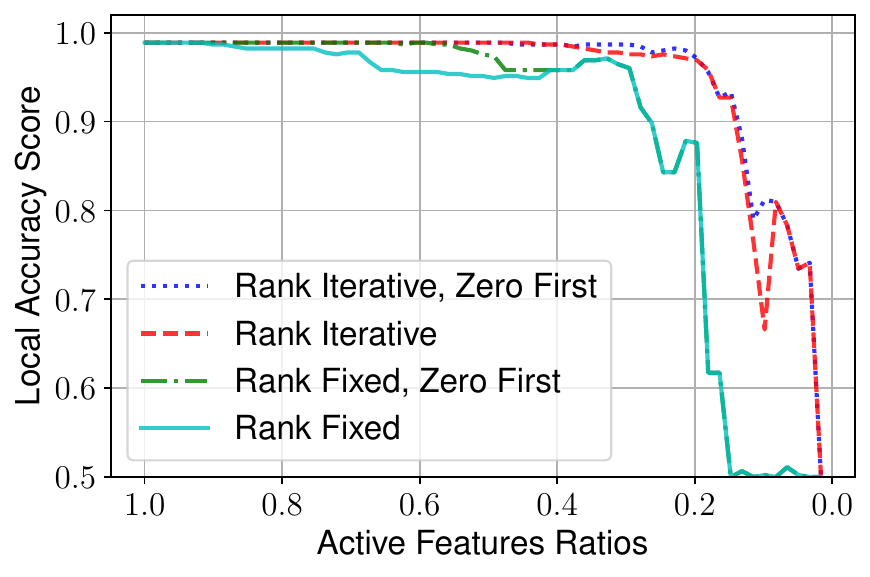}
    \caption{\acl{rfe} techniques with different feature ranking and removal conditions.}
    \label{fig:rank}
\end{figure}

In \Cref{fig:rank}, we display the outcome of \ac{rfe} under many conditions. Specifically, we investigate the differences between using a pre-computed ranking versus an iterative method that recalculates each feature's rank after removing the least effective one. Moreover, the least effective feature is determined by prioritizing non-influential ones first (i.e.,~those with zero influence), or selecting the feature with the lowest rank, which could potentially be negative, if its removal benefits the model.
The most substantial differences arise when recomputing the rank at each iteration rather than removing zero-ranked features first. We notice differences between the two sets of lines in \Cref{fig:rank}. Recomputing the ranking enables the model to consistently evaluate the best local solution based on the current state, thereby preserving performance while decreasing the number of features. On the other hand, using a fixed rank results in an earlier accuracy drop when using more features.

\begin{figure}[tb]
    \centering
    \includegraphics[width=\columnwidth]{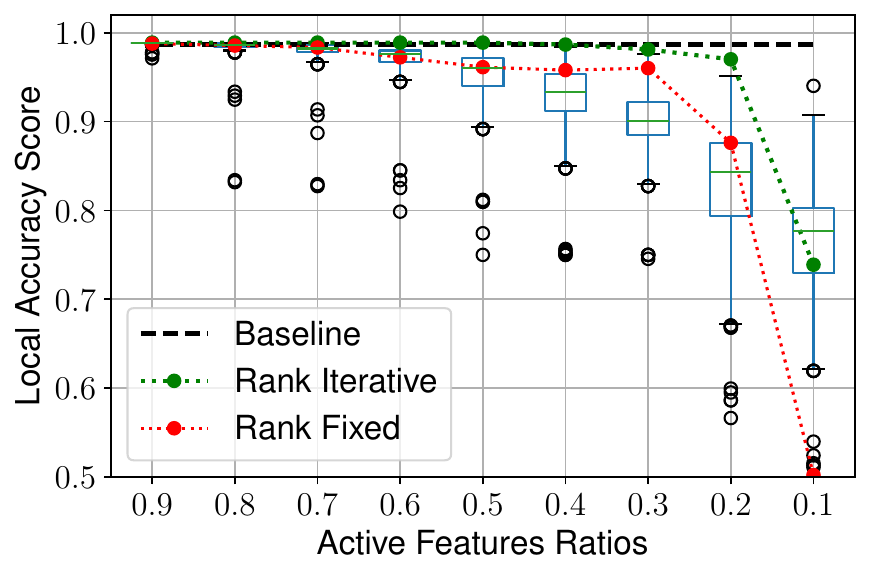}
    \caption{Feature subset \acl{ss} results versus the average of \acl{rfe} techniques.}
    \label{fig:onlysubset}
\end{figure}

Based on the iterative findings, we compute the average between the fixed and iterative rank results and use these values as a baseline. We report the outcomes of the \acl{ss} in \Cref{fig:onlysubset} alongside the model's baseline (\ie using the full feature set with a ratio $1.0$).
For each ratio of active features (x-axis), we focus on the most relevant results, excluding instances where the model's ability to detect attacks deteriorated (\ie all predictions are zeros or ones). This criterion requires achieving an accuracy greater than $0.5$ in every traffic category, namely \texttt{BENIGN} and \texttt{DDoS}.
The results in \Cref{fig:onlysubset} suggest that, on average, given the \ac{bigml} model under test, there are alternative combinations of active features that offer comparable, if not better, outcomes. Especially when decreasing the ratio of active features, most results fall below the previously computed iterative baselines, yet there is a margin for improvements. While the performance remains similar to the baseline for most ratios, for very small subsets such as $0.1$, the algorithm identifies alternative and improved solutions.

\subsection{Model Pruning}

\begin{figure}[tb]
    \centering
    \includegraphics[width=\columnwidth]{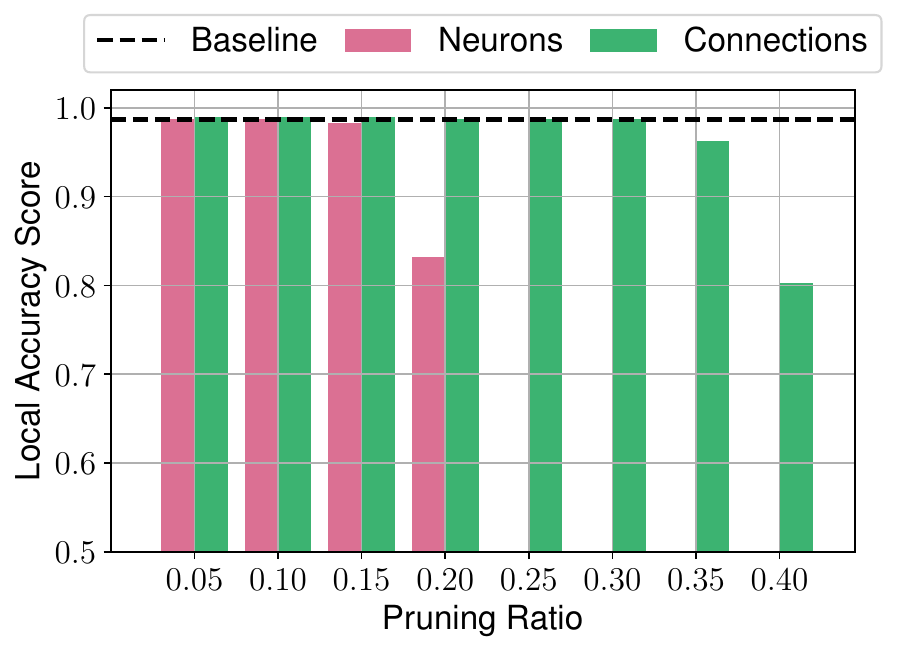}
    \caption{Comparison of model pruning techniques when removing neurons versus connections.}
    \label{fig:onlyprune}
\end{figure}

\Cref{fig:onlyprune} shows the accuracy score difference between pruned versions of the original \ac{bigml} by removing neurons or connections.
Removing neurons preserves the performance of the initial model until approximately $15\%$ of its internal neurons are removed (\ie pruning ratio $0.15$). At a ratio of $0.20$, the model starts degrading, achieving an acceptable value of accuracy of $0.82$: beyond this threshold, the algorithm fails to identify a viable model. On the other hand, pruning connections allow for a more in-depth pruning of the model, up to $35\%$ with near-optimal performance and $40\%$ with still high accuracy (\ie $0.8$) of all internal network links. However, after this value, this approach also fails. 
For the upcoming test with only model pruning, we will use a pruning ratio of $0.15$ and neuron removal as the reference method, allowing us to simulate the requirements for a lighter setup. Despite only a small portion of the model being pruned, the resulting model is lighter and capable of faster predictions.

\begin{table}[tb]
    \centering
    \caption{Neurons-pruned models performance metrics.}
    \label{tab:onlyprune}
    \begin{tabular}{ccc}
\toprule
Prune Ratio & Inference Time (ns) & Memory (KB) \\
\midrule
0.00 & 307 & 323 \\
0.05 & 301 & 288 \\
0.10 & 293 & 263 \\
0.15 & 289 & 239 \\
0.20 & 284 & 216 \\
0.25 & 270 & 194 \\
\vdots & \vdots & \vdots\\
0.95 & 193 & 4 \\

\bottomrule
\end{tabular}
\end{table}

\begin{figure}[!t]
    \centering
    \includegraphics[width=\columnwidth]{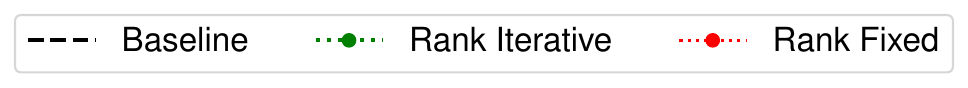}\\
    \includegraphics[width=\columnwidth]{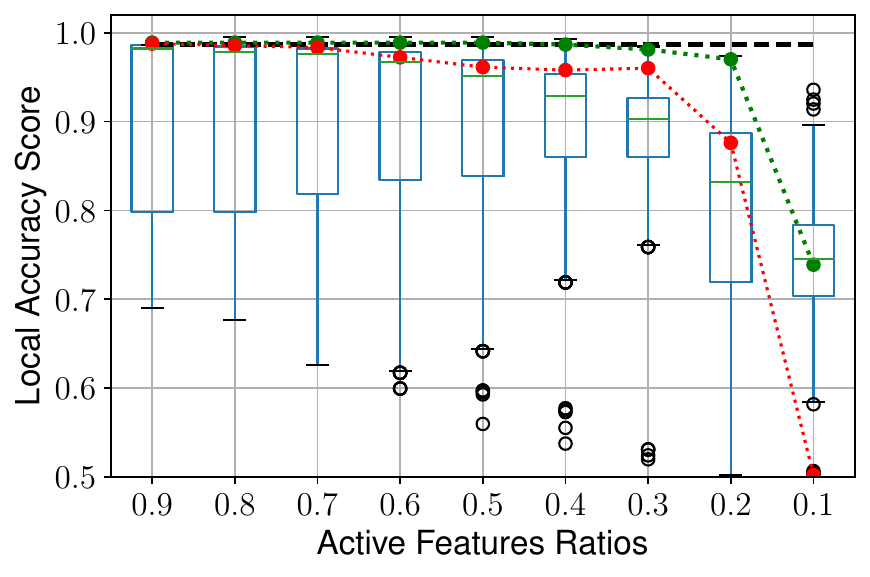}
    \caption{Combining model pruning with feature subset search while reducing the active features ratios. The highest pruning ratio achieved is $0.25$ for $x<0.8$, and $0.2$ otherwise.}
    \label{fig:combo}
\end{figure}

In \Cref{tab:onlyprune}, we report the average inference time and memory usage of the models pruned at different rates, omitting some values in the range for readability. We observe that by removing just 15\% of neurons, nearly one-third of RAM usage for model execution can be saved. Our analysis excludes the potential use of auxiliary and support variables, which could introduce implementation-specific considerations. While the impact on inference speed may not appear significant with neuron removal, in scenarios involving constrained or real-time devices handling many samples per second, the difference becomes relevant. With a pruning ratio of $0.15$ that still preserves the model quality, we observed an almost $6\%$ increase in inference speed (\ie \SI{289}{\nano\second} against \SI{307}{\nano\second} of the original model).

\subsection{Feature Selection and Model Pruning}

The results of combining model pruning with feature subset search methods are depicted in \Cref{fig:combo}. 

\begin{figure*}[t]
    \centering
    \includegraphics[width=\textwidth]{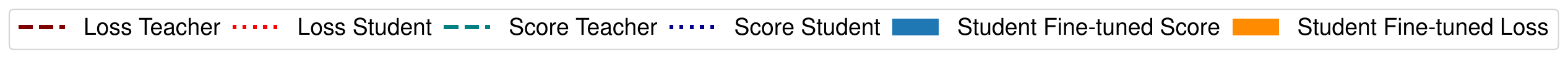}\\
    \begin{subfigure}{0.49\textwidth}
    \includegraphics[width=\textwidth]{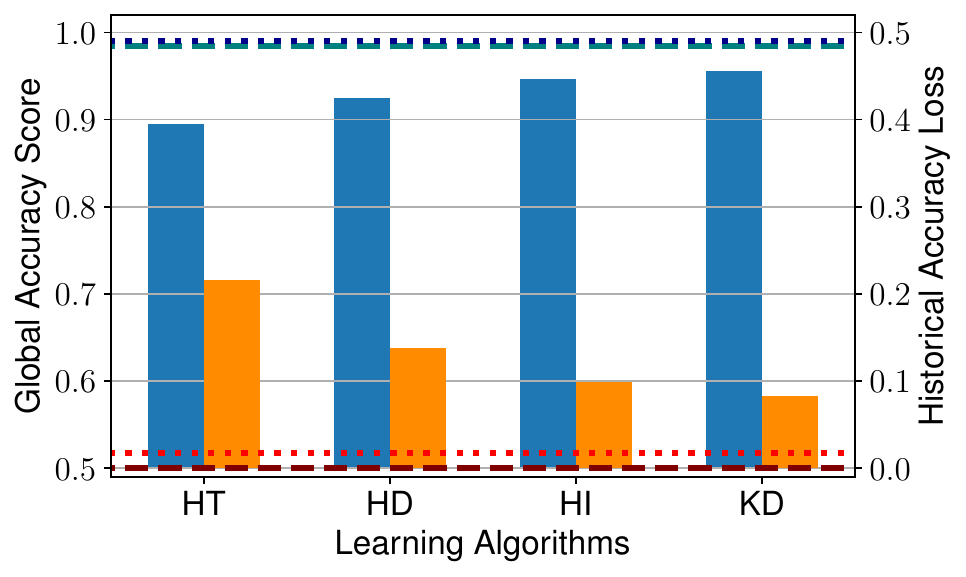}
    \caption{Student model is a \ac{bigml}.}
    \label{fig:a}
    \end{subfigure}
    \begin{subfigure}{0.49\textwidth}
    \includegraphics[width=\textwidth]{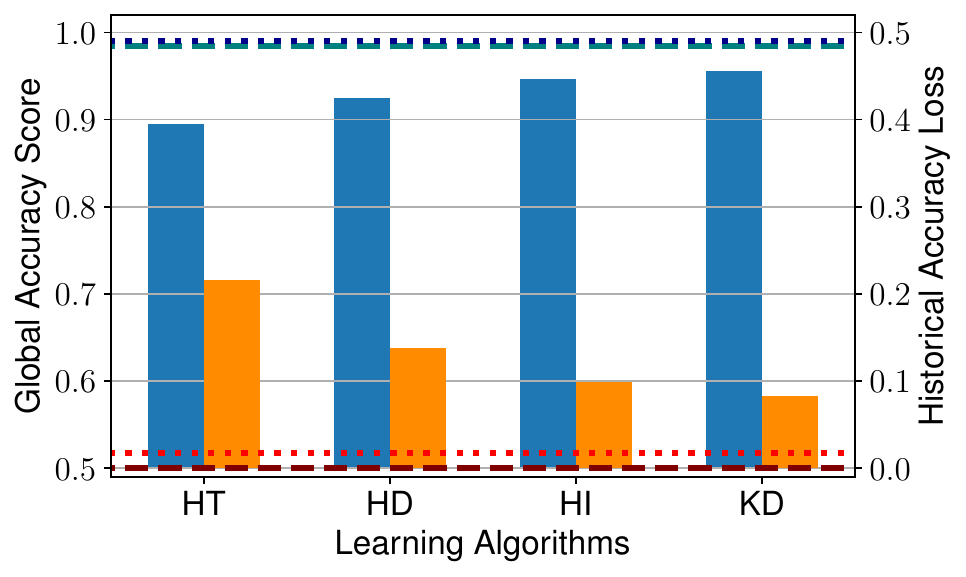}
    \caption{Student model is a \acf{prunedbigml}.}
    \label{fig:b}
    \end{subfigure}
    \caption{Global accuracy score and loss on historical data of the student model when fine-tuning on local traffic.}
    \label{fig:testbigmlandprune}
\end{figure*}

We highlight that the maximum pruning ratio achieved during this process is $0.25$ for all active feature ratios, except for $0.9$ and $0.8$, where it is $0.2$, suggesting a more limited scope for neuron pruning.
\Cref{fig:combo} suggests that combining model pruning and feature stochastic search processes can still yield nearly optimal results, occasionally surpassing previously computed baselines. Compared to the \ac{ss} alone in \Cref{fig:onlysubset}, this plot displays more varied quantiles, particularly for lower ratios of active features, suggesting a more diverse outcome. Nevertheless, the results remain highly comparable, and beyond a ratio of $0.5$, it is even feasible to exceed the original \ac{bigml} baseline.

\subsection{Fine-tuning with Full Feature Set}

We leverage the \ac{bigml} model as the teacher model, the ``oracle'' that, when required by the learning algorithm, guides the fine-tuning of the derived models. The student model under tests is sequentially one of those presented in \Cref{tab:matrixmodels}. The following plots provide for both the teacher and student model \textit{(i)} the baseline global accuracy; the \textit{(ii)} baseline accuracy loss concerning the historical knowledge; the \textit{(iii)} global accuracy of the student model at the end of the fine-tuning (\ie student fine-tuned), along with the respective historical loss (\ie blue and orange bars respectively).
The x-axis shows the different learning algorithms introduced in \Cref{tab:learningalgorithms}.

\Cref{fig:testbigmlandprune} presents the results of the test on centralized deployments presented in \Cref{tab:matrixmodels}. In particular, in \Cref{fig:a} a replica of the initial \ac{bigml} serves as a student model, while in \Cref{fig:b} we use the pruned counterpart \ac{prunedbigml} with a ratio of $0.15$, resulting in a lighter architecture.

In this experiment, we observe how the learning algorithm impacts the model's ability to preserve its historical knowledge. \Cref{fig:a} suggests that the most problematic algorithm used in those circumstances is the standard \ac{hardtrue}, which leverages the true data labels. Not only does the resulting fine-tuned student model achieve a lower overall accuracy score, but it also suffers from a significant loss of historical knowledge. On the other hand, the other algorithms preserve both the overall model performance and its previous knowledge. From the plot, we can conclude that \ac{kd}, which uses only the soft label predicted from the teacher model, is a suitable candidate for fine-tuning, as the student model retains both historical and local knowledge.

When removing neurons from the initial model as in \Cref{fig:b}, the student model exhibits a non-zero historical loss baseline before fine-tuning, which then increases. While all the learning algorithms struggle to retain historical knowledge, there are differences among them. The less detrimental algorithm is \ac{kd}, achieving the highest accuracy score while keeping the knowledge loss below $10\%$. Moreover, the \ac{hardinferred} performs relatively well in this context, nearly matching \ac{kd} performance. This can be attributed to the teacher's predictions, which may not always be precise ($0.99$ of accuracy), enabling the \ac{prunedbigml} model to learn from these inaccuracies. Finally, an algorithm such as \ac{hardtrue} likely pushes the model training in the opposite direction in such cases, increasing the performance gap compared to the baselines. 

\subsection{Fine-tuning with Partial Feature Set}

\begin{figure*}[tb]
    \centering
    \includegraphics[width=\textwidth]{res/plots/legend2.pdf}\\
    \begin{tabular}{cccc}
    &
    \underline{Active Features Ratio 0.3}
    &
    \underline{Active Features Ratio 0.5}
    &
    \underline{Active Features Ratio 0.8}
    \\
         \rotatebox{90}{\underline{Case 1}}
         &
        \raisebox{-0.5\height}{\includegraphics[width=.3\textwidth,trim={0 0 0.8cm 0}]{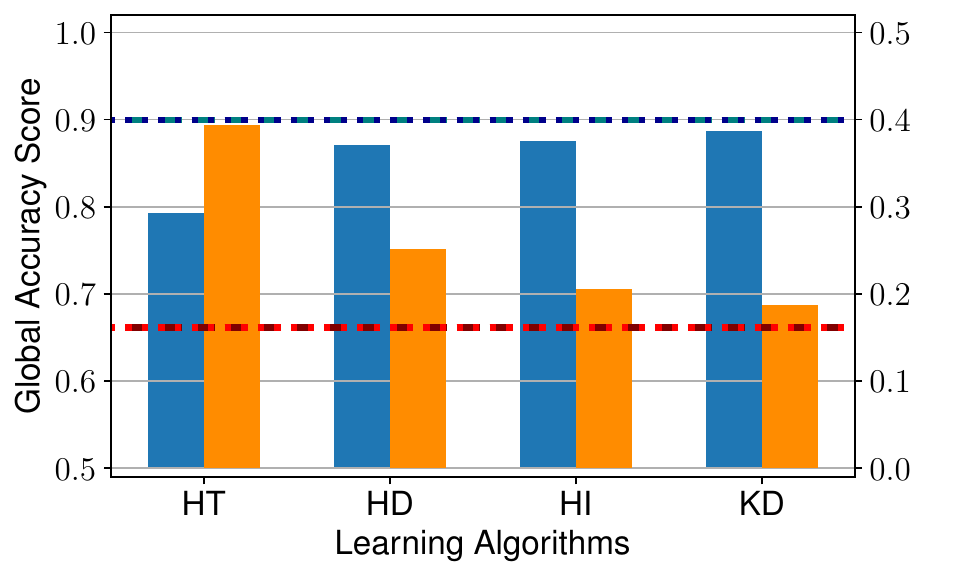}}
        &
        \raisebox{-0.5\height}{\includegraphics[width=.3\textwidth,trim={0.4cm 0 0.4cm 0}]{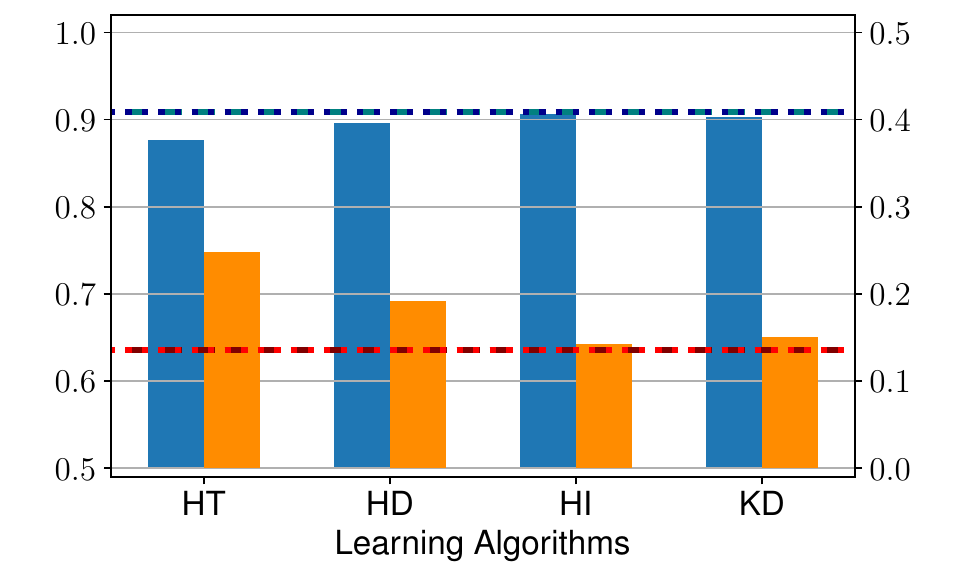}}
        &
        \raisebox{-0.5\height}{\includegraphics[width=.3\textwidth,trim={0.8cm 0 0 0}]{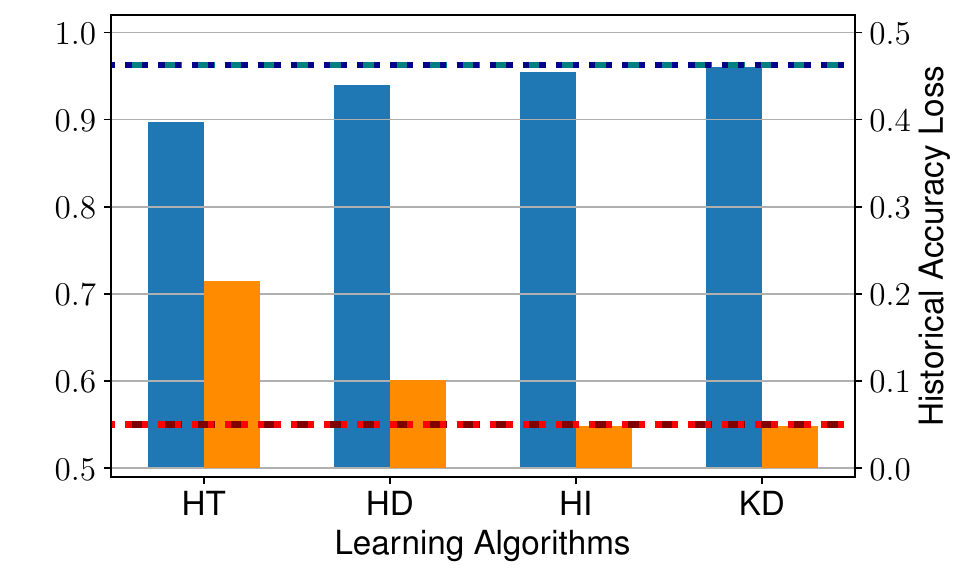}}
        \\
        
        \rotatebox{90}{\underline{Case 2}}
         &
        \raisebox{-0.5\height}{\includegraphics[width=.3\textwidth,trim={0 0 0.8cm 0}]{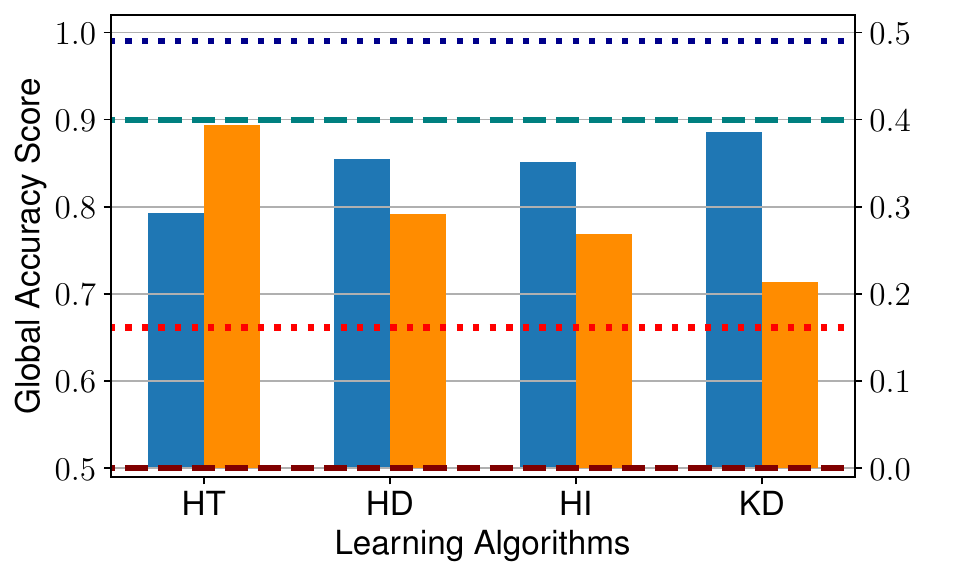}}
        &
        \raisebox{-0.5\height}{\includegraphics[width=.3\textwidth,trim={0.4cm 0 0.4cm 0}]{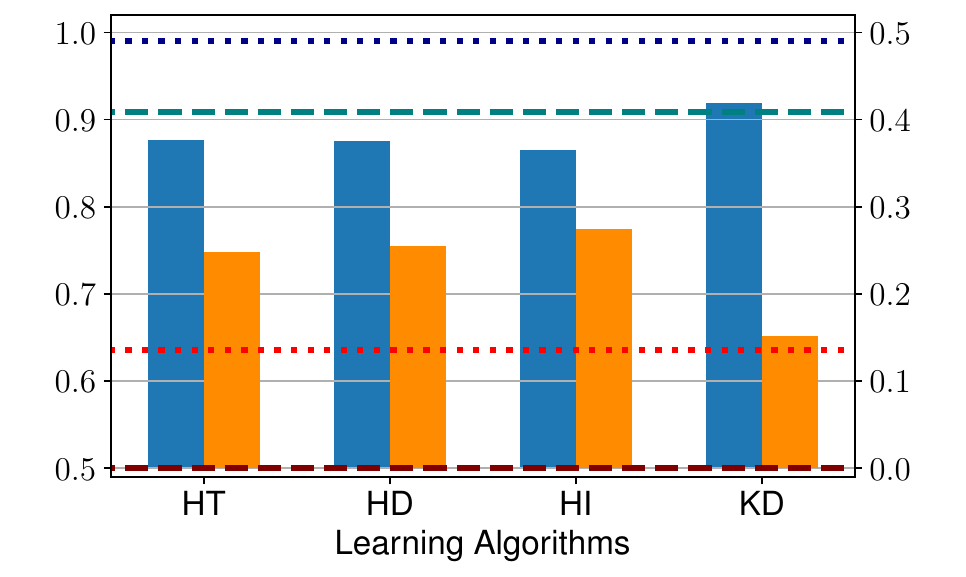}}
        &
        \raisebox{-0.5\height}{\includegraphics[width=.3\textwidth,trim={0.8cm 0 0 0}]{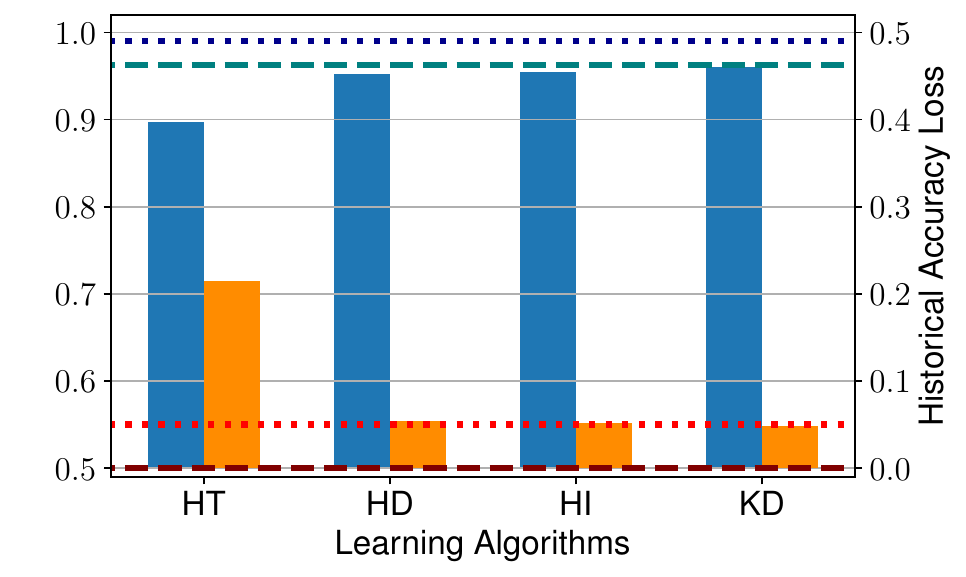}}
        \\

        \rotatebox{90}{\underline{Case 3}}
         &
        \raisebox{-0.5\height}{\includegraphics[width=.3\textwidth,trim={0 0 0.8cm 0}]{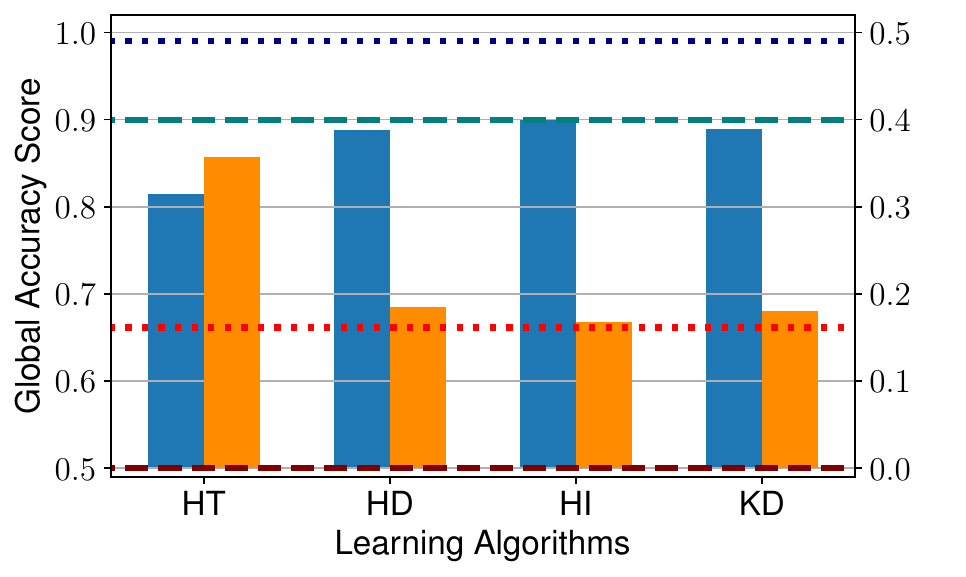}}
        &
        \raisebox{-0.5\height}{\includegraphics[width=.3\textwidth,trim={0.4cm 0 0.4cm 0}]{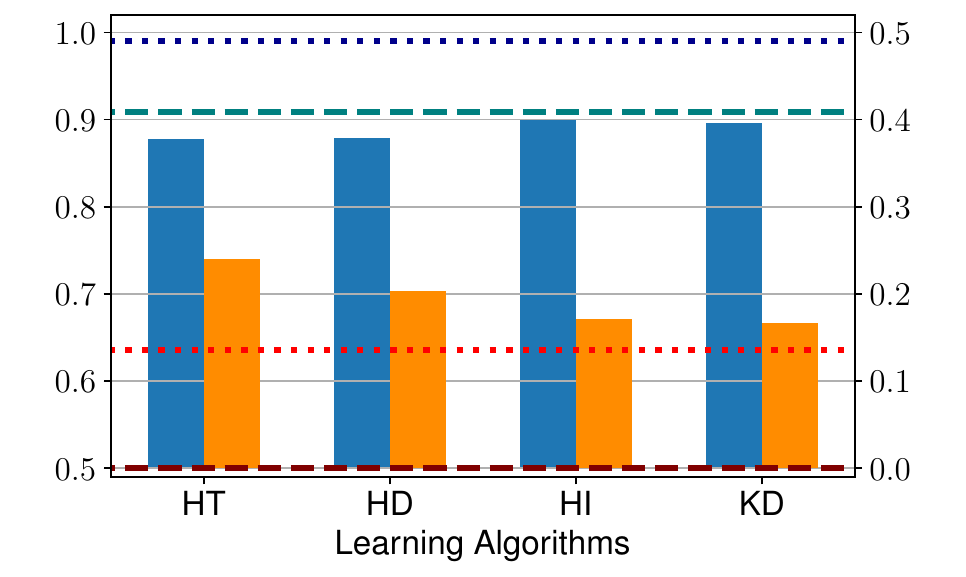}}
        &
        \raisebox{-0.5\height}{\includegraphics[width=.3\textwidth,trim={0.8cm 0 0 0}]{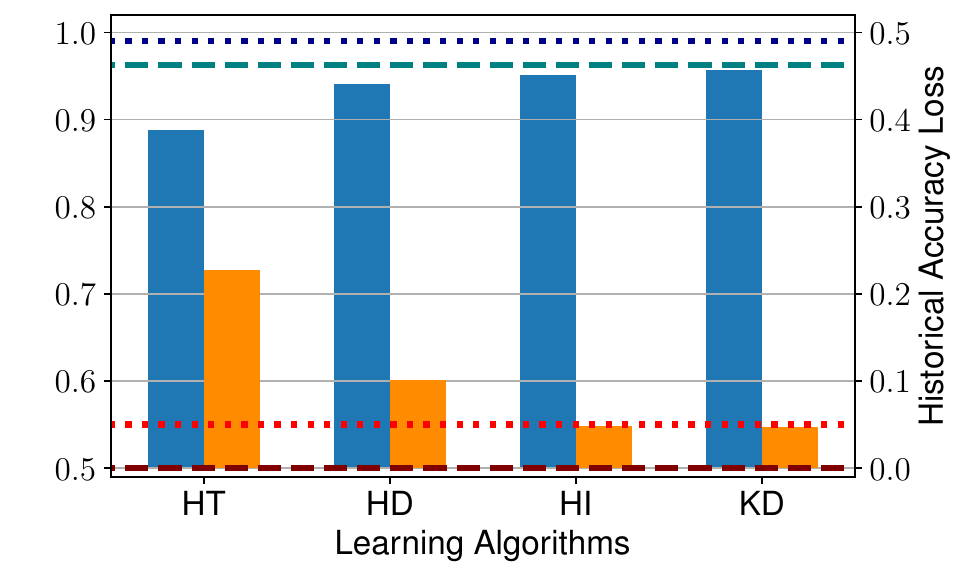}}
        \\

        \rotatebox{90}{\underline{Case 4}}
         &
        \raisebox{-0.5\height}{\includegraphics[width=.3\textwidth,trim={0 0 0.8cm 0}]{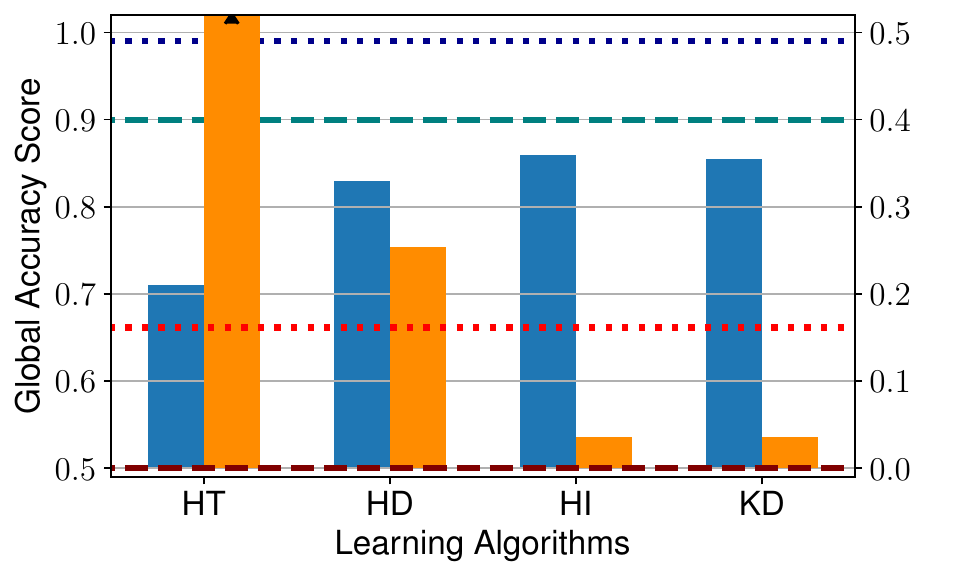}}
        &
        \raisebox{-0.5\height}{\includegraphics[width=.3\textwidth,trim={0.4cm 0 0.4cm 0}]{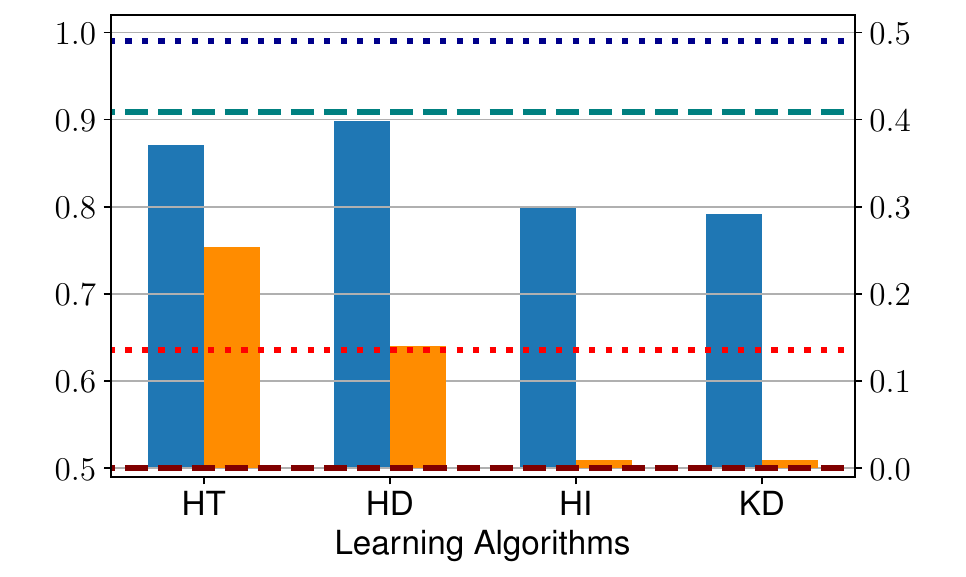}}
        &
        \raisebox{-0.5\height}{\includegraphics[width=.3\textwidth,trim={0.8cm 0 0 0}]{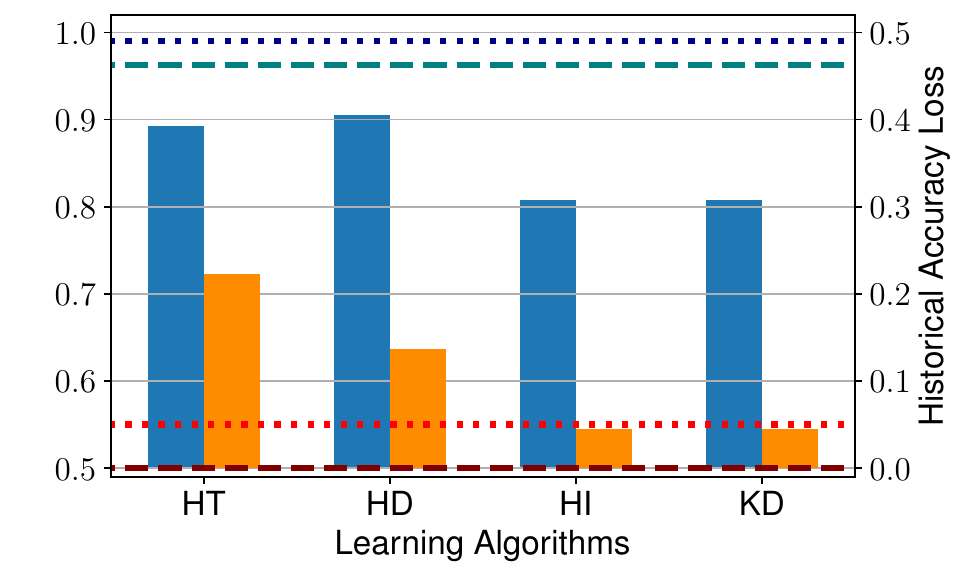}}
        \\
    \end{tabular}
        \caption{Global accuracy score (blue) and loss on historical data (orange) of the student model when fine-tuning on local traffic. The student is a replica of the original \ac{bigml} model but provided with smaller feature sets.}
    \label{fig:testedgebigml}
\end{figure*}

We now examine the outcomes of fine-tuning models in edge conditions, simulating missing features from the input data. As previously outlined in \Cref{tab:matrixscenarios}, we identify many scenarios for such deployment.
The first model we evaluate in these circumstances is \ac{edgebigml}, a replica of \ac{bigml} but provided with fewer input features, assuming that an edge device can manage the complexity of the complete model. The results are presented in \Cref{fig:testedgebigml}. For each case (\ie rows in the plots), we present results for active features ratios (\ie columns in the plot) of $0.3$, $0.5$, and $0.8$.

In the \texttt{Case 1}, we observe a decrease in the baselines for global accuracy scores, from $0.99$ in the previous cases (\eg \Cref{fig:testbigmlandprune}) to $0.90$ when using a smaller feature subset (\ie first two columns of plots). At the same time, the baselines for historical accuracy loss increased to nearly $0.15$, except when using a high ratio of active features such as $0.8$, which allows both the teacher and the student models to start with higher performance. In these experiments, \ac{hardtrue} and \ac{avgtruesoftinferred} cause more historical loss and struggle to compete with the other algorithms in achieving global accuracy.
Once again, the \ac{kd} proved to be the most performant algorithm, with a small performance degradation and almost entirely preserving the baseline performance before the fine-tuning.

In \texttt{Case 2}, where the teacher has access to the features, its baseline returns to the previous near-optimal values. Decreasing the ratio of active features results in increasing historical loss for the student model, especially with algorithms that use the original sample labels (\ie \ac{avgtruesoftinferred} and \ac{hardtrue}). Interestingly, at an active feature ratio of $0.5$ (\ie middle column), only \ac{kd} manages to surpass the student model threshold in performance while preserving historical loss. 

In \texttt{Case 3}, the student model receives the teacher data during the learning process, consisting of the same samples as the student but with the full feature set. The plots suggest that this condition benefits nearly all learning algorithms, except \ac{hardtrue}, which leads to a remarkable increase in the historical loss, despite achieving a high level of global accuracy. In the case of an active feature ratio of $0.3$, all algorithms managed to preserve the model's baseline, except for \ac{hardtrue}. In the previous case (\ie case 2), we observe that this was not possible, and the model experienced greater loss with any algorithm. Moving to a higher active feature ratio of $0.5$ and $0.8$, we observe no significant differences compared to the previous scenario.

In \texttt{Case 4}, starting from the last column with the active feature ratio of $0.8$, we note that algorithms such as \ac{kd} and \ac{hardinferred} manage to preserve historical loss, but fail to improve the model's global accuracy. On the other hand, \ac{avgtruesoftinferred} performs better, achieving high accuracy while keeping the historical loss around $0.1$, although it cannot reach the baseline. Despite presenting the highest loss in the historical data, \ac{hardtrue} scores higher than other algorithms with an accuracy of nearly $0.9$, due to the mixed data usage during learning, which enables the model to efficiently learn $DDoS$ but especially \texttt{BENIGN} traffic patterns, which still comprise $50\%$ of the dataset across all categories. Moving to a smaller subset of $0.5$, \ac{avgtruesoftinferred} is the most performant algorithm, while \ac{kd} can almost nullify the historical loss. 
% On the other hand, 
However,
when using a subset of $0.3$, the situation reverses. \ac{kd} can keep the accuracy score close to the baseline while reducing historical loss to less than $0.05$, similar to \ac{hardinferred}.

\begin{figure*}[tb]
    \centering
    \includegraphics[width=\textwidth]{res/plots/legend2.pdf}\\
    \begin{tabular}{cccc}
    &
    \underline{Active Features Ratio 0.3}
    &
    \underline{Active Features Ratio 0.5}
    &
    \underline{Active Features Ratio 0.8}
    \\
         \rotatebox{90}{\underline{Case 1}}
         &
        \raisebox{-0.5\height}{\includegraphics[width=.3\textwidth,trim={0 0 0.8cm 0}]{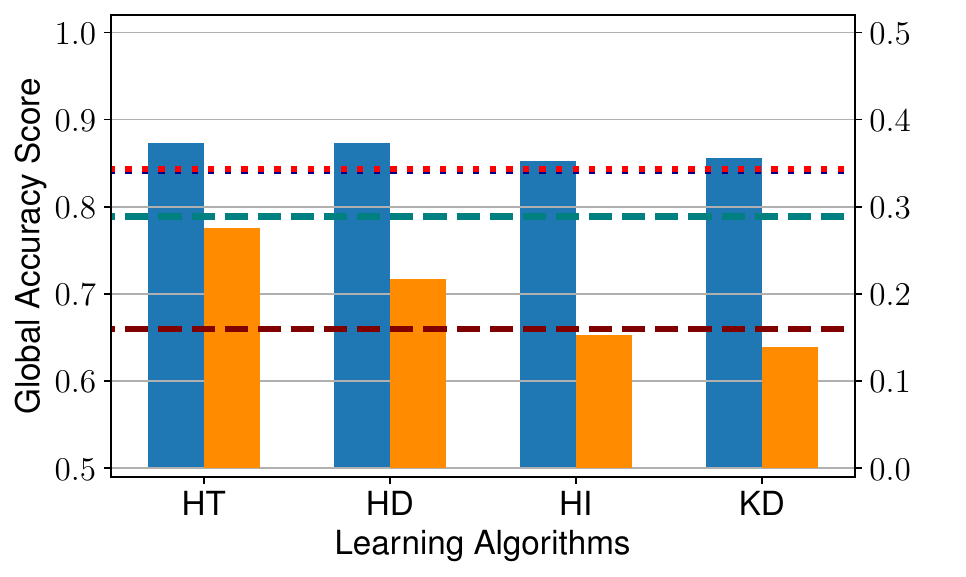}}
        &
        \raisebox{-0.5\height}{\includegraphics[width=.3\textwidth,trim={0.4cm 0 0.4cm 0}]{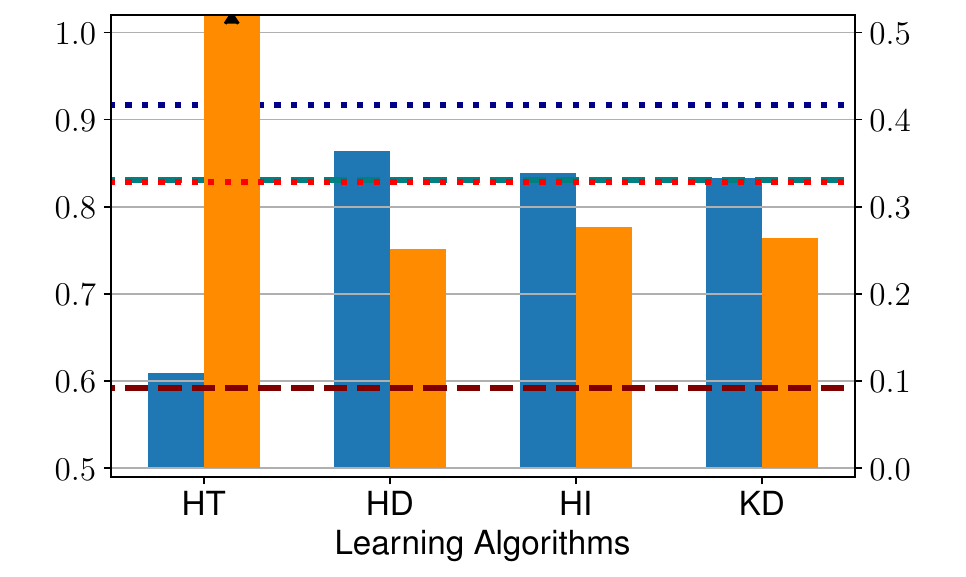}}
        &
        \raisebox{-0.5\height}{\includegraphics[width=.3\textwidth,trim={0.8cm 0 0 0}]{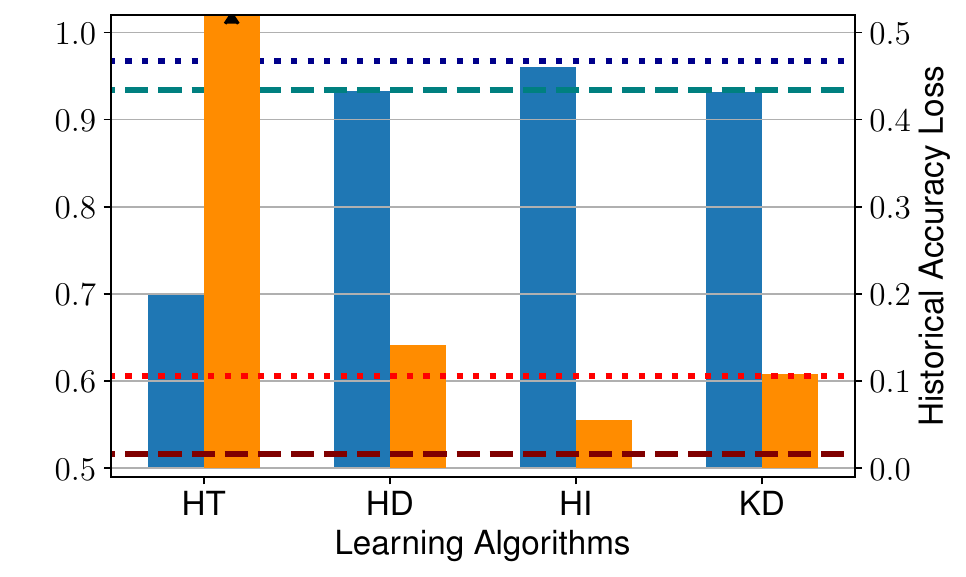}}
        \\
        
        \rotatebox{90}{\underline{Case 2}}
         &
        \raisebox{-0.5\height}{\includegraphics[width=.3\textwidth,trim={0 0 0.8cm 0}]{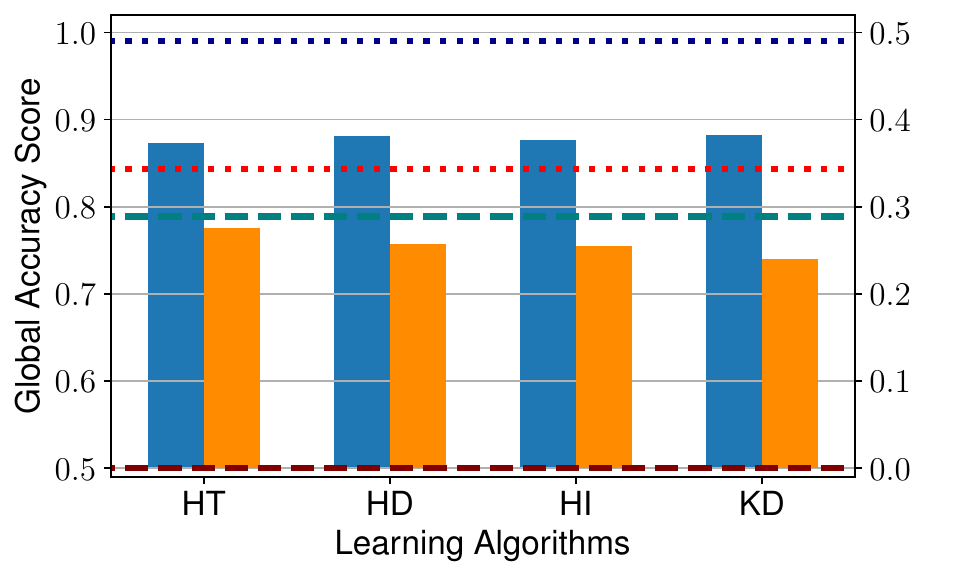}}
        &
        \raisebox{-0.5\height}{\includegraphics[width=.3\textwidth,trim={0.4cm 0 0.4cm 0}]{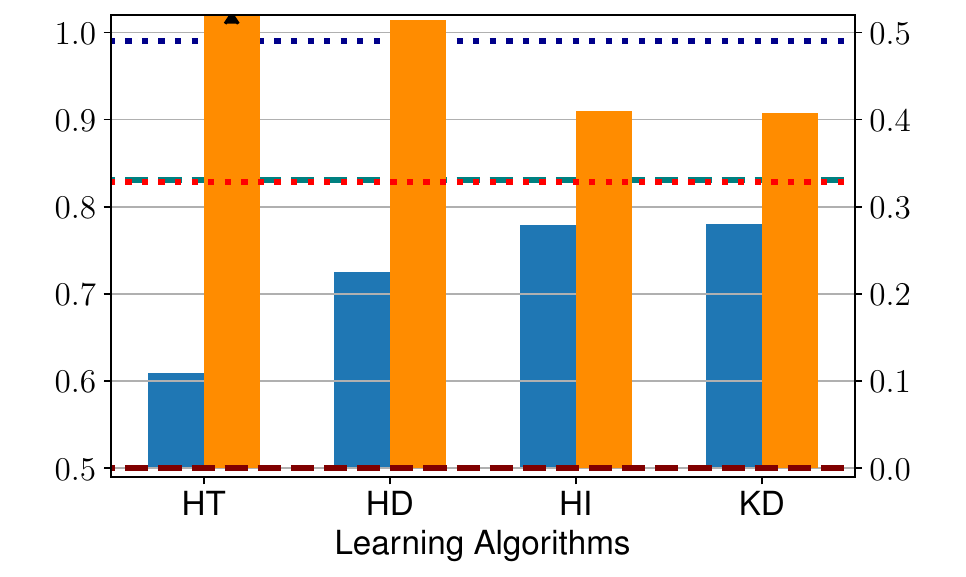}}
        &
        \raisebox{-0.5\height}{\includegraphics[width=.3\textwidth,trim={0.8cm 0 0 0}]{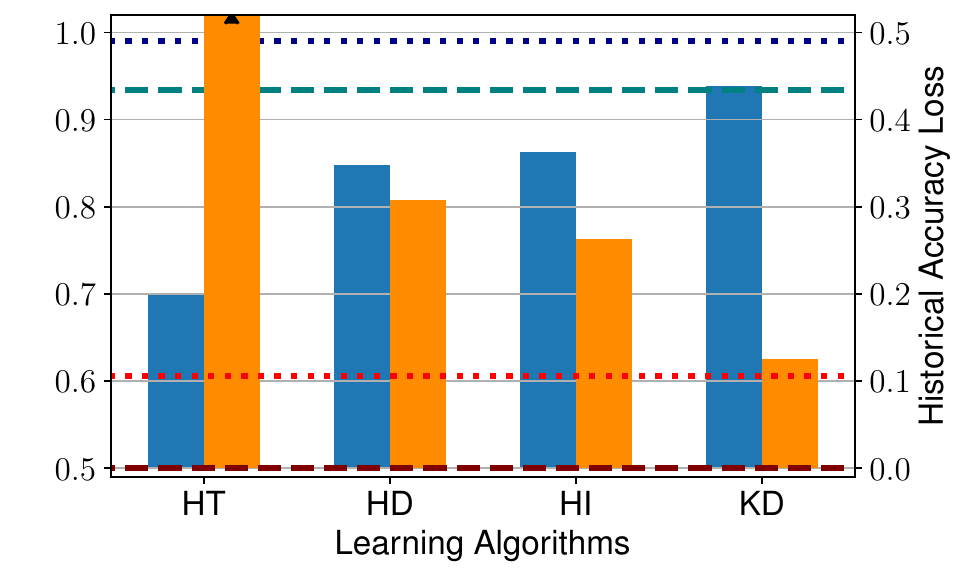}}
        \\

        \rotatebox{90}{\underline{Case 3}}
         &
        \raisebox{-0.5\height}{\includegraphics[width=.3\textwidth,trim={0 0 0.8cm 0}]{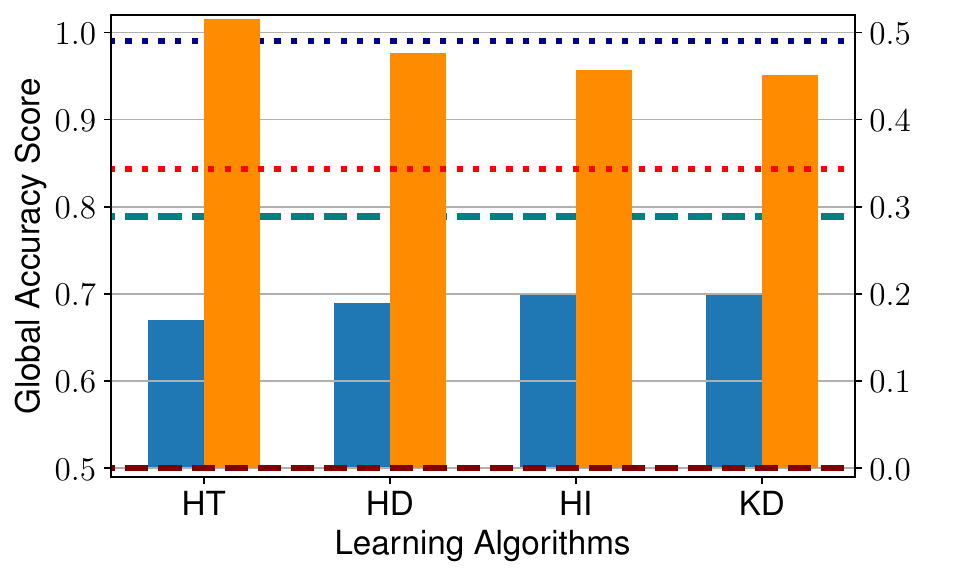}}
        &
        \raisebox{-0.5\height}{\includegraphics[width=.3\textwidth,trim={0.4cm 0 0.4cm 0}]{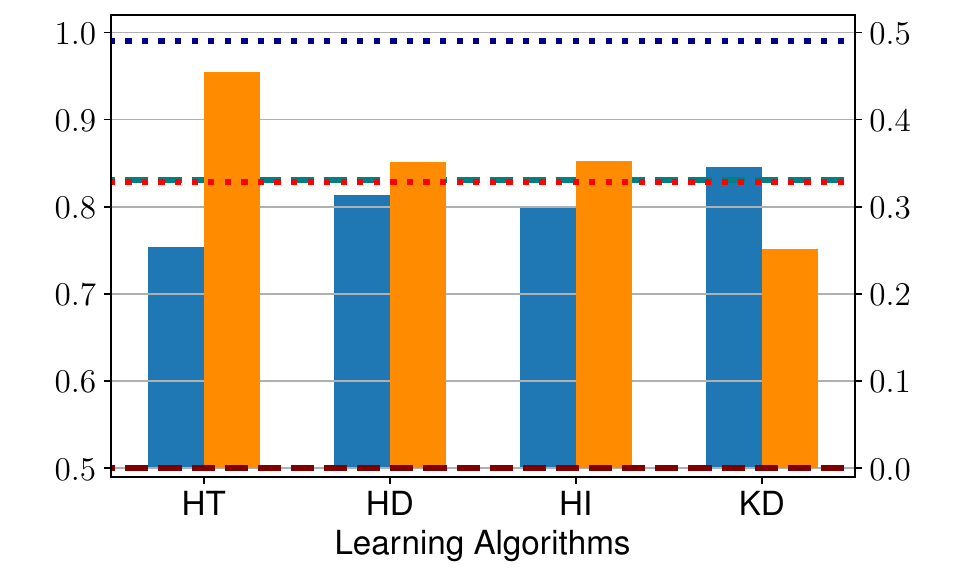}}
        &
        \raisebox{-0.5\height}{\includegraphics[width=.3\textwidth,trim={0.8cm 0 0 0}]{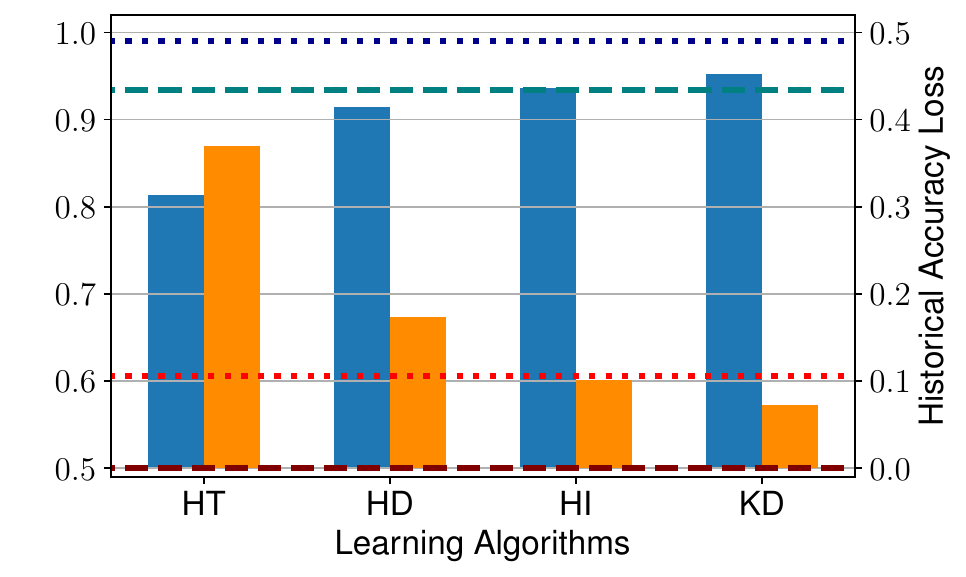}}
        \\

        \rotatebox{90}{\underline{Case 4}}
         &
        \raisebox{-0.5\height}{\includegraphics[width=.3\textwidth,trim={0 0 0.8cm 0}]{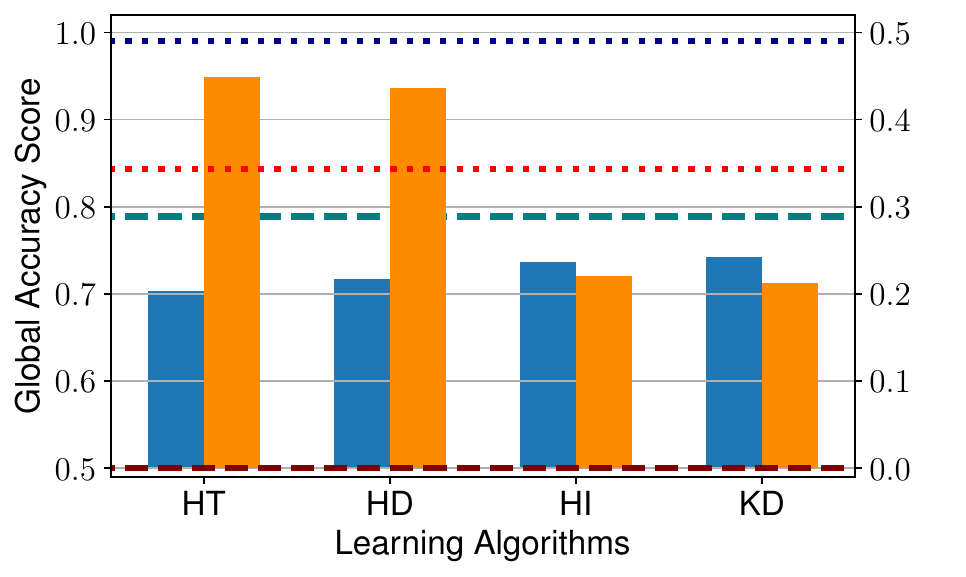}}
        &
        \raisebox{-0.5\height}{\includegraphics[width=.3\textwidth,trim={0.4cm 0 0.4cm 0}]{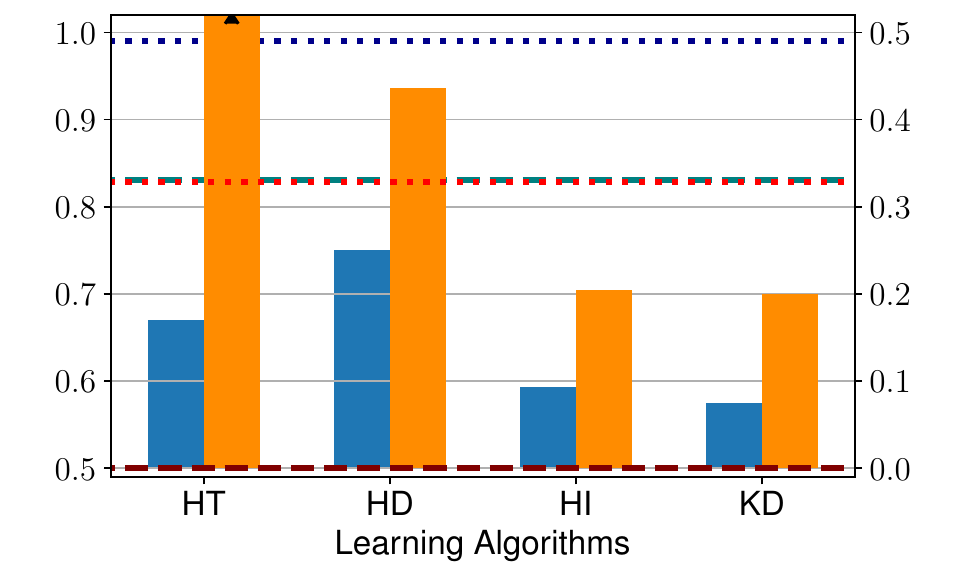}}
        &
        \raisebox{-0.5\height}{\includegraphics[width=.3\textwidth,trim={0.8cm 0 0 0}]{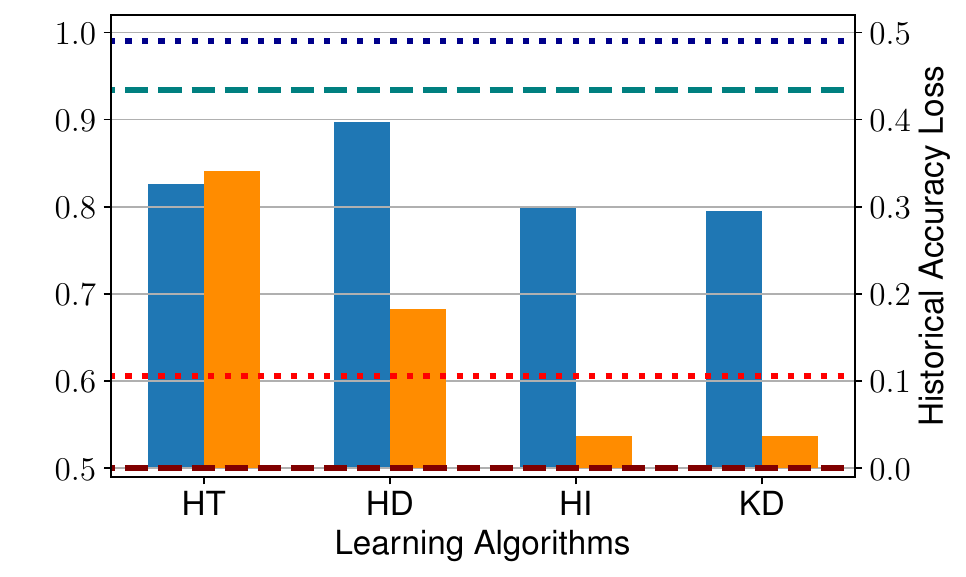}}
        \\
    \end{tabular}
    \caption{Global accuracy score (blue) and loss on historical data (orange) of the student model when fine-tuning on local traffic. The student is obtained by combining model pruning with feature subset search, and is provided with partial feature sets.}
    \label{fig:testedgesmallmlprunefirst}
\end{figure*}

We now analyze the most challenging scenario, where both model pruning and feature selection techniques are applied. We refer to this type of deployment as \ac{edgeprunedbigml}, and we present its results in \Cref{fig:testedgesmallmlprunefirst}.
In \texttt{Case 1}, with a small active feature ratio of $0.3$, every learning algorithm enables surpassing the teacher accuracy baseline. In addition, \ac{kd} and \ac{hardinferred} lower the historical loss to a minimum of $0.13$, slightly below the teacher model. On the other hand, algorithms such as \ac{hardtrue} cannot preserve the same level of knowledge, but they do help increase global accuracy, further specializing the model for local traffic. As the active feature ratio increases, \ac{hardtrue} struggles to preserve both the global accuracy and historical loss, making it unsuitable for scenarios where the teacher model lacks access to the complete feature set. In \texttt{Case 2}, a subset ratio of $0.3$ fails to identify a single outstanding algorithm better than the others, as their results are very similar. 
% On the other hand, 
Conversely,
in \texttt{Case 3}, when learning using teacher data, a subset ratio of $0.3$ does not yield a valid model that preserves its original performance. Increasing the ratio to $0.5$ and $0.8$, \ac{kd} and \ac{hardinferred} emerge as the best algorithms, enhancing overall accuracy and reducing historical loss. Finally, using a mixed input as in \texttt{Case 4} does not allow for achieving good performance, regardless of the subset ratio. While \ac{kd} can reduce historical loss compared to the baseline, the achieved accuracy score is still significantly lower than the baseline.

In summary, we evaluated various combinations of student models with the same \ac{bigml} teacher model obtained during the initial phase of collaborative training. When fine-tuning a model, there is no rule of thumb to identify in advance the better solution for the given configuration. In most of our cases, the \ac{kd} is the preferred choice, followed by \ac{hardinferred} and \ac{avgtruesoftinferred}. In border case scenarios with both a portion of the model missing and fewer input features, \ac{hardtrue} can better fine-tune the student model to the local traffic at the cost of losing more historical knowledge. On the other hand, the \ac{kd} can preserve historical knowledge in almost every scenario.
\section{Conclusion} \label{sec:conclusion}

In this paper, we have presented \ac{name}, an approach to tackle the adaptability problem of \ac{ml}-based \acp{ids} to the specific target device needs, enabling deployment in peripheral devices with constrained resources.
\ac{name} defines a pipeline that starts with a pre-trained base reference model resulting from collaborative training across many organizations, such as using \ac{fl}. The pipeline integrates feature selection, model pruning, and fine-tuning techniques from local data sources to explore sub-optimal subsets of the original feature set alongside a pruned version of the reference model, allowing the identification of less resource-demanding configurations for the overall \ac{ids}. To the best of our knowledge, no prior work has integrated these tasks into a single activity.

We have demonstrated the effectiveness of \ac{name} to deliver sub-optimal solutions in different setups with incrementally challenging conditions, simulating the transition from the core to the edge of an infrastructure. Additionally, many fine-tuning algorithms have been tested to identify optimal methods for distilling knowledge from the original \ac{fl}-trained model to the lighter setups, proving that Knowledge Distillation with soft labels reduces the effects of catastrophic forgetting in almost every testing scenario.

\iffalse
\section*{Declaration of competing interest}
The authors declare that they have no known competing financial
interests or personal relationships that could have appeared to influence the work reported in this paper.

\section*{Data availability}
The data used for the experiments is public and referenced in the
paper.
\fi
\section*{Acknowledgement}
This work was partially supported by the European Union’s Horizon Europe Programme under grant agreement No 101070473 (project
FLUIDOS).

\bibliographystyle{IEEEtran} 
\bibliography{bibliography}

\end{document}